\documentclass{aa}  

\usepackage{graphicx}
\usepackage{txfonts}
\usepackage{natbib}
\usepackage{placeins}
\usepackage{hyperref}
\usepackage{graphicx}
\usepackage{amssymb}
\usepackage{amsmath}
\usepackage{xcolor}
\usepackage{subcaption}
\usepackage{longtable}
\usepackage{statmath}

\usepackage{siunitx}

\newcommand{\um}{\SI{}{\micro\meter}}

\newcommand\jwst{\textit{JWST}}
\newcommand\hst{\textit{HST}}

\def\arcmin{\ifmmode {^{\prime}}\else $^{\prime}$\fi}
\def\arcsec{\ifmmode {^{\prime\prime}}\else $^{\prime\prime}$\fi}
\newcommand\qth{$^{\rm th}$}
\newcommand{\av}{A$_{V}$}
\newcommand{\orcid}[1]{\includegraphics[scale=0.06]{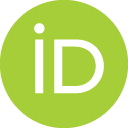} \href{https://orcid.org/#1}{#1}}

\newcommand{\rev}[1]{{#1}}

\begin{document} 

  \title{The Structure of Massive Star-Forming Galaxies from JWST and ALMA: Dusty, High Redshift Disk Galaxies}
  \author{Steven Gillman,\inst{1,2}\thanks{\href{orcids}{ORCIDs} listed on final page}
           Ian Smail, \inst{3}
           Bitten Gullberg, \inst{1,2}
           A. M. Swinbank, \inst{3}
           Aswin P. Vijayan, \inst{1,2,4}
           Minju Lee, \inst{1,2}
           Gabe Brammer,  \inst{1,5}
           U. Dudzevi\v{c}i\={u}t\.{e}, \inst{6}
           Thomas R. Greve,  \inst{1,2,7} 
           Omar Almaini, \inst{8}
           Malte Brinch, \inst{1,2}
           Scott C. Chapman, \inst{9}
           Chian-Chou Chen, \inst{10}
           Soh Ikarashi, \inst{11,12,13}
           Yuichi Matsuda, \inst{11,14,15}
           Wei-Hao Wang, \inst{10}
           Fabian Walter, \inst{6,16}
           \and Paul P. van der Werf\,\inst{17}}
           
   \institute{Cosmic Dawn Center (DAWN), Denmark\\
    \email{srigi@space.dtu.dk}
    \and
        DTU-Space, Elektrovej, Building 328 , 2800, Kgs. Lyngby, Denmark
    \and 
        Centre for Extragalactic Astronomy, Department of Physics, Durham University, South Road, Durham DH1 3LE, UK
    \and 
        Astronomy Centre, University of Sussex, Falmer, Brighton BN1 9QH, UK
    \and
        Niels Bohr Institute, University of Copenhagen, Jagtvej 128, DK-2200 Copenhagen N, Denmark
    \and
        Max-Planck-Institut für Astronomie, Königstuhl 17, 69117 Heidelberg, Germany
    \and 
        Dept. of Physics and Astronomy, University College London, Gower Street, London WC1E 6BT, United Kingdom
     \and 
        School of Physics and Astronomy, University of Nottingham, Nottingham NG7 2RD, UK
    \and 
         Dept of physics, Dalhousie University, Halifax NS, Canada 
    \and 
         Academia Sinica Institute of Astronomy and Astrophysics (ASIAA), No. 1, Section 4, Roosevelt Road, Taipei 10617
    \and 
        National Astronomical Observatory of Japan, 2-21-1 Osawa, Mitaka, Tokyo, 181-8588, Japan
    \and 
        Junior College, Fukuoka Institute of Technology, 3-30-1 Wajiro-higashi, Higashi-ku, Fukuoka, 811-0295 Japan
    \and 
        Department of Physics, General Studies, College of Engineering, Nihon University, 1 Nakagawara, Tokusada, Tamuramachi, Koriyama, Fukushima, 963-8642, Japan  
    \and Graduate University for Advanced Studies (SOKENDAI), Osawa 2-21-1, Mitaka, Tokyo 181-8588, Japan
    \and 
        Cahill Center for Astronomy and Astrophysics, California Institute of Technology, MS 249-17, Pasadena, CA 91125, USA
    \and
        National Radio Astronomy Observatory, Pete V. Domenici Array Science Center, P.O. Box O, Socorro, NM 87801, USA    
    \and 
        Leiden Observatory, Leiden University, P.O. Box 9513, 2300 RA Leiden, The Netherlands}    
    \date{Accepted September 13, 2024}
  
  \abstract{We present an analysis of the \jwst\ NIRCam and MIRI morphological and structural properties of 80 massive ($\log_{10}(M_\ast[M_{\odot}])$\,=\,11.2\,$\pm$\,0.1) dusty star-forming galaxies at $z$\,$=$\,2.7$^{+1.2}_{-0.7}$, identified as sub-millimetre galaxies (SMGs) by ALMA,  that have been observed as part of the \jwst{} PRIMER project. To compare the structure of these massive, active galaxies to more typical less actively star-forming galaxies, we define \rev{two comparison samples. The first of 850 field galaxies matched in specific star-formation rate and redshift and the second of 80 field galaxies matched in stellar mass}. From visual classification of the SMGs, we identify 20\,$\pm$\,5\% as candidate late-stage major mergers, a further 40\,$\pm$\,10\% as potential minor mergers and 40\,$\pm$\,10\% which have comparatively undisturbed disk-like morphologies, with no obvious massive neighbours on $\lesssim$\,20\,--\,30\,kpc (projected) scales. These rates are comparable to those for the field samples and indicate that the majority of the sub-millimetre-detected galaxies are not late-stage major mergers, but have interaction rates similar to the general field population at $z$\,$\sim$\,2--3.  Through a multi-wavelength morphological analysis, using parametric and non-parametric techniques, we establish that SMGs have comparable near-infrared, mass normalised, sizes to the less active population, $\rm R_{50}^{F444W}$\,=\,2.7\,$\pm$\,0.2\,kpc versus $\rm R_{50}^{F444W}$\,=\,3.1\,$\pm$\,0.1\,kpc, but exhibit lower S\'ersic indices, consistent with bulge-less disks: $n_{\rm F444W}$\,=\,1.1\,$\pm$\,0.1, compared to $n_{\rm F444W}$\,=\,1.9\,$\pm$\,0.1 for the less active field \rev{and $n_{\rm F444W}$\,=\,2.8\,$\pm$\,0.2 for the most massive field galaxies}.  The SMGs exhibit greater single-S\'ersic fit residuals and their morphologies are more structured at 2\um\ relative to 4\um\ when compared to the field galaxies.  This appears to be caused by  significant structured dust content in the SMGs and we find evidence for dust reddening as the origin of the morphological differences by identifying a strong  correlation between the F200W$-$F444W pixel colour  and the 870\um\ surface brightness using high-resolution ALMA observations. We conclude that SMGs and \rev{both massive} and less massive star-forming galaxies at the same epochs share a common disk-like structure, but the weaker bulge components (and potentially lower black hole masses) of the SMGs result in their gas disks being less stable. Consequently, the combination of high gas masses and instabilities triggered either secularly or by minor external perturbations results in higher levels of activity (and dust content) in SMGs compared to typical star-forming galaxies.}
  
\keywords{Galaxies: high-redshift --
             Galaxies: structure --
             Galaxies: evolution -- 
             Sub-millimetre: galaxies}
\titlerunning{A Combined JWST and ALMA View of the Structure of Sub-millimetre Galaxies}
\authorrunning{S. Gillman et al.}
\maketitle
%
%________________________________________________________________

\section{Introduction}

The relative proportions of highly-dust-obscured and less-obscured star formation appears to vary over the history of the Universe, with the dust-obscured component dominating in galaxy populations at $z$\,$\sim$\,4--5 down to the present day \citep[e.g.,][]{Dunlop2017,Bouwens2020,Long2023}.  The cause of this transition in star-formation mode may reflect  the growing metallicity of the interstellar medium (ISM) in galaxies, less efficient removal of dust from deeper potential wells or structural or geometrical changes in the star-forming regions within galaxies.    

The most extreme examples at high redshifts of systems dominated by  dust-obscured  star formation are the sub-millimetre galaxies (SMGs) with dust masses of $M_{\rm d}$\,$\sim$\,10$^{8-9}$\,M$_\odot$ and far-infrared luminosities of $L_{\rm IR}$\,$\sim$\,10$^{12-13}$\,L$_\odot$ \citep[e.g.,][]{Magnelli2012,Rowlands2014,Miettinen2017b,Dud2020}, placing them into the Ultra or Hyperluminous Infrared Galaxies (U/HyLIRGs) classes, see  \citet{Hodge2020} for a full review.  Most of these galaxies show high star-formation rates, SFR\,$\sim$\,10$^{2-3}$\,M$_{\odot}$\,yr$^{-1}$\citep[e.g.,][]{Swinbank2014}, and correspondingly short gas consumption timescales \citep[e.g.,][]{Greve2005,Miettinen2017a,Tacconi2018,Birkin2021}, suggesting that they represent relatively short-lived starbursts, $\lesssim$\,100\,Myrs,  which will result in massive systems with  stellar masses of $M_\ast$\,$\sim$\,10$^{11}$\,M$_{\odot}$ \citep[e.g.,][]{Wardlow2011,Simpson2014,Miettinen2017a,Dud2020}.

The high star-formation rates and large stellar masses of SMGs have proved challenging to reproduce in theoretical galaxy formation models \citep[e.g.,][]{Baugh2005,Swinbank2008,Hayward2013,Mcalpine2019}, although more recent attempts have been more successful \citep[e.g.,][]{Lovell2022,Lower2022,Cochrane2023}.  The models suggest that the high star-formation rates in this population are driven by a mix of secular instabilities in gas-rich disks and dynamical triggers due to minor and major mergers \citep[e.g.,][]{Mcalpine2019}.   However, attempts to observationally test these claims using the available  rest-frame ultra-violet (UV) \textit{Hubble Space Telescope} (\hst) imaging of $z$\,$\sim$\,1--3 SMGs were challenging \citep[e.g.,][]{Chapman2003}. While the galaxies frequently exhibited irregular morphologies with  apparently multiple components \citep[e.g.,][]{Swinbank2010,Chen2015,Gomez2018,Zavala2018,Cowie2018,Lang2019,Ling2022}, their significant dust attenuation   ($A_{V} \gtrsim $\,2--6, e.g., \citealt{Dud2020}) means that robustly assessing the intrinsic stellar mass morphology of the galaxies from these data was incredibly difficult. This is especially the case for the subset of the population that is undetectable in the near-infrared (``NIR-faint'' SMGs, \citealt{Simpson2014,Dud2020,Smail2021,Ikarashi2022,Kokorev2023}).  Detecting and resolving the rest-frame near-infrared emission of SMGs, which is much less affected by dust attenuation and more closely traces the bulk of the stellar population, is essential to robustly constrain the morphology of  their stellar mass. However, this has been beyond the reach of  previous instrumentation such as the Infrared Array Camera (IRAC; \citealt{Fazio2004}) on the Spitzer Space Telescope, which,  whilst providing the required near-infrared wavelength coverage from 3.6\,--\,8\um, lacked the sub-arcsecond spatial resolution required to constrain the underlying morphologies of the stellar emission in SMGs. 

However, some progress has been made in understanding the structure of SMGs using high-resolution observations of the dust continuum and gas kinematics in SMGs \citep[e.g.,][]{Simpson2015,Ikarashi2015,Fujimoto2017}.   Thus FWHM\,$\sim$\,0\farcs{1}--0\farcs{2}  sub-millimetre  mapping with ALMA has revealed bright dust continuum emission arising from compact disk-like structures  (i.e., $R_{\rm e}$\,$\sim$\,1\,--\,2\,kpc, S\'ersic $n$\,$\approx$\,1) as well as evidence for a fainter more extended component with $R_{\rm e}$\,$\simeq$\,4\,kpc \citep[e.g.,][]{Gullberg2019,Ivison2020}. The bright compact component is roughly half the size of the galaxy extent in \hst\ $H$-band imaging  \citep{Hodge2016,Gullberg2019,Chen2020,Cochrane2021}.  At the highest resolutions, FWHM\,$\sim$\,$0\farcs{05}$, \citet{Hodge2019} identified potential arms and bar-like structures in the bright dust continuum from a small sample of high-redshift SMGs. 

Resolved kinematic studies using molecular and atomic fine structure emission lines in the rest-frame far-infrared and sub-/millimetre have uncovered disk-like kinematics for at least a significant fraction of the population \citep[e.g.,][]{Hodge2012,Chen2017,Lelli2021,Rizzo2021,Amvrosiadis2023}. With the advent of the James Webb Space Telescope (JWST, \citealt{Gardner2023}), high-resolution observations from the near (1\um) to the mid-infrared (25\um) are now feasible, enabling the infrared emission from counterparts of sub-millimetre bright galaxies to be resolved and the issue of their stellar morphology and structures to be finally quantified. 

Initial morphological studies  with \jwst\ have found  size variations with wavelength in sub-millimetre galaxies and less active colour-selected populations at $z$\,$\sim$\,2 \citep{Chen2022,Cheng2023a,Suess2022},  with a  reduction in half-light radius when observed in the  rest-frame near-infrared compared to the rest-frame optical/UV as previously studied with \hst . A number of studies have subsequently expanded the samples with  \jwst\ and ALMA coverage, confirming that  SMGs become more compact (higher concentration, smaller size) at longer wavelengths \citep[e.g.][]{Cheng2023a,Gillman2023,Price2023} with the potential impact of dust on the observed morphology becomingly less evident at longer wavelengths \citep[e.g.,][]{Kokorev2022,Wu2023,Kam2023,Sun2024}. There are also claims of frequent stellar bars in many high-redshift star-forming galaxies \citep{Guo2023,Mckinney2024}, including examples in SMGs \citep{Smail2023}. However, all of these studies  suffer from modest sample statistics, limiting the interpretations of their findings.

In this paper, we present an analysis of the structural properties of a statistically robust sample of ALMA-detected, sub-millimetre detected galaxies that have  been observed with \jwst\ NIRCam and MIRI. In Section \ref{sec:sample} we define the sample of SMGs that we use in our analysis whilst in Section \ref{sec:data} we present the observations, data reduction and analysis undertaken on these systems (and matched control samples of less active galaxies selected from the field). In Section \ref{Sec:Results} \rev{we give our results and in Section \ref{Sec:Disc} we discuss their implications} before summarising our main conclusions in Section \ref{Sec:Conc}. Throughout the paper, we assume a $\Lambda$CDM cosmology with $\Omega_{\rm m} = 0.3$, $\Omega_{\Lambda} = 0.7$, and $H_0 = 70\,\mathrm{km\,s^{-1}\,Mpc^{-1}}$. All quoted magnitudes are on the AB system and stellar masses are calculated assuming a Chabrier initial mass function (IMF) \citep{Chabrier2003}. 

\begin{figure*}
    \includegraphics[width=\linewidth]{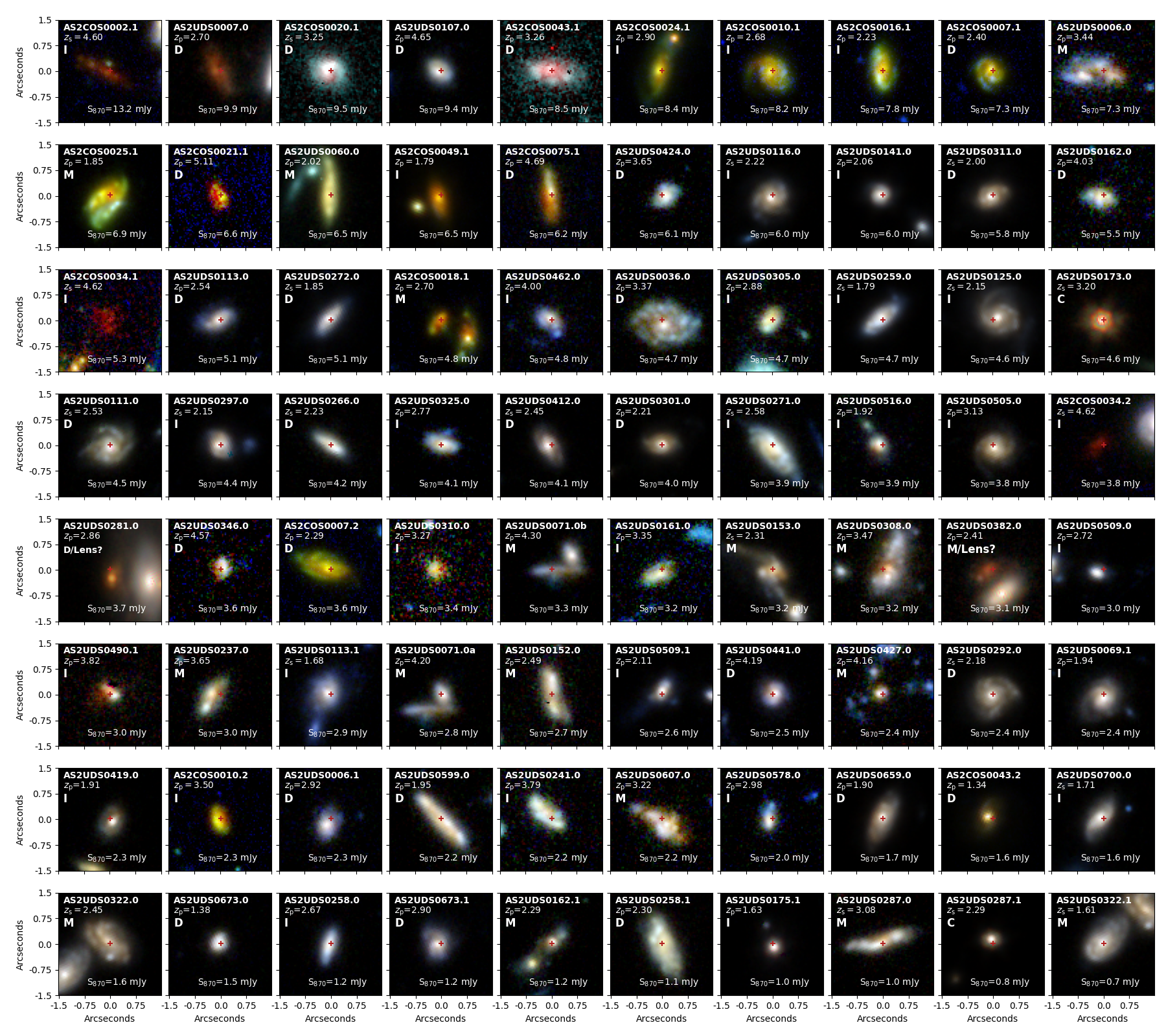}
    \caption{The combined AS2UDS and AS2COSMOS SMG sample in the PRIMER survey ranked in descending ALMA 870\um\ flux. For each SMG we show the 3\farcs{0}$\,\times\,$3\farcs{0} false colour image (F444W/F356W/F277W as R/G/B), labelled with the SMG ID, redshift ({\sc magphys} derived, $z_{\rm p}$, or spectroscopic, $z_{\rm s}$) and ALMA 870\um\ flux density. We further label the SMGs visually classified as major mergers (M) \rev{minor mergers or interactions (I),  regular disks (D) or compact (C) as well as potentially those} affected by gravitational lensing (Lens). The red cross indicates the ALMA 870\um\ position.}
    \label{Fig:Cimg}
\end{figure*}

%sample selection
\section{Sample selection}\label{sec:sample}

%AS2UDS and AS2COSMOS summary
To build a sample of ALMA-detected SMGs with near-infrared \jwst{} coverage, we utilise the AS2UDS \citep{Stach2019} and AS2COSMOS \citep{Simpson2020} surveys as our parent sample. The AS2UDS and AS2COSMOS surveys are ALMA 870\um\ follow-up programs of 716 and 180 850\um\ SCUBA-2 sources that are detected at $>$\,4$\sigma$ in the S2CLS \citep{Geach2017} map of the UKIDSS Ultra-Deep Survey (UDS; \citealt{Lawrence2007})  and the S2COSMOS \citep{Simpson2019}  map of the Cosmic Evolution Survey (COSMOS; \citealt{Scoville2007})  respectively. These two surveys provide an initial candidate sample of 896 SMGs with precise ALMA identifications, selected at S/N\,$\gtrsim$\,4, for which we can use to characterise their near-infrared \jwst{} counterparts.

%PRIMER
We  cross-match the ALMA positions for the SMGs with the footprint of the \jwst/NIRCam and \jwst/MIRI observations from The Public Release IMaging for Extragalactic Research (PRIMER; \citealt{Dunlop2021}) survey. PRIMER is a multi-band, multi-instrument survey of UDS and COSMOS covering 234 and 144 sq. arcmin respectively. The observations consist of the seven wide-band and one medium-band NIRCam filter (F090W, F115W, F150W, F200W, F277W, F356W, F410M, F444W) as well as two wide-band MIRI filters (F770W, F1800W). Parts of the UDS and COSMOS fields also benefit from \hst\ ACS and WFC3 coverage from CANDELS \citep{Grogin2011} sampling the observed frame optical (0.4\um) to near-infrared (1.6\um).

%cross matching
For our analysis, we require the SMGs to be covered by the \jwst/NIRCam long-wavelength filter F444W ensuring that the observed-frame near-infrared is sampled. This results in a sample of 66 AS2UDS and 22 AS2COSMOS SMGs. A summary of the ALMA properties and \hst\ and \jwst\ coverage is presented in Appendix A. Of the 88 SMGs, 12 have no \hst{} coverage. Two SMGs have \jwst{} imaging in only three bands (F444W, F277W, F200W) while the other 86 SMGs are covered by three or more \jwst{} bands, with 37 SMGs having the maximum ten available \jwst\ filters covering from 0.9 to 18\,\um\, (Appendix A).

The 88 SMGs have a median ALMA 870\um\ flux of $S_{\rm 870\um}$\,=\,3.8\,$\pm$\,0.4\,mJy, with a 16\qth\ to 84\qth\ percentile range of $S_{\rm 870\um}$\,=1.8\,--\,6.1\,mJy with the brightest, AS2COS0002.1 having $S_{\rm 870\um}$\,=\,13.2\,mJy (see Appendix A). The SMGs sample the bulk of the AS2COSMOS and AS2UDS $S_{\rm 870\um}$ distribution which ranges from $S_{\rm 870\um}$\,=\,0.6\,--\,19.2\,mJy. In addition the majority of the AS2UDS SMGs (51/66) were observed with ALMA at resolutions of $\simeq$\,0\farcs{3} FWHM as presented in \citet{Stach2019} with 15 SMGs observed at a higher resolution of $\simeq$\,0\farcs{2} FWHM \citep[see][]{Gullberg2019}. The ALMA maps of the 22 SMGs in the COSMOS field  were tapered to a resolution of 0\farcs{8} FWHM \citep{Simpson2020}.

\section{Reduction and analysis}\label{sec:data}

In this section, we present our data reduction and compilation of the photometric data that we use to characterise the multi-wavelength properties of the SMGs and their analysis. 

\subsection{\jwst\ and \hst}\label{subsec:HST_JWST}

% Reduction
We homogeneously process the \jwst\  NIRCam and MIRI observations, retrieving the Level-2 data products (version\,=\,{jwst\_1069.pmap}) from the STScI website\footnote{ https://mast.stsci.edu/} and processing them with the {\sc{grizli}} pipeline \citep{Brammer2021, Brammer2022}\footnote{For full details of the reduction process see: \url{https://dawn-cph.github.io/dja/imaging/v7/}}. For the NIRCam data, additional steps were employed to deal with diagonal striping seen in some exposures, cosmic rays and stray light. Whilst for the MIRI exposures time-dependent sky flats are applied in an approach similar to recent \jwst\ studies \citep[e.g.,][]{Yang2023}. We further incorporate the available optical and near-infrared data available in the Complete Hubble Archive for Galaxy Evolution \citep[CHArGE,][]{Kokorev2022}, providing multi-band imaging from 0.4 to 1.6\um\ for a subset of the sources (see Appendix A). We align all the imaging to Gaia DR3 \citep{Gaia2021}, co-add, and drizzle the final mosaics to a $0\farcs04$ pixel scale \citep{Fruchter2002} for all \jwst\ and \hst\ filters. 

\begin{figure*}
    \centering
    \includegraphics[width=\linewidth,trim={0.1cm 0cm 0cm 0cm},clip]{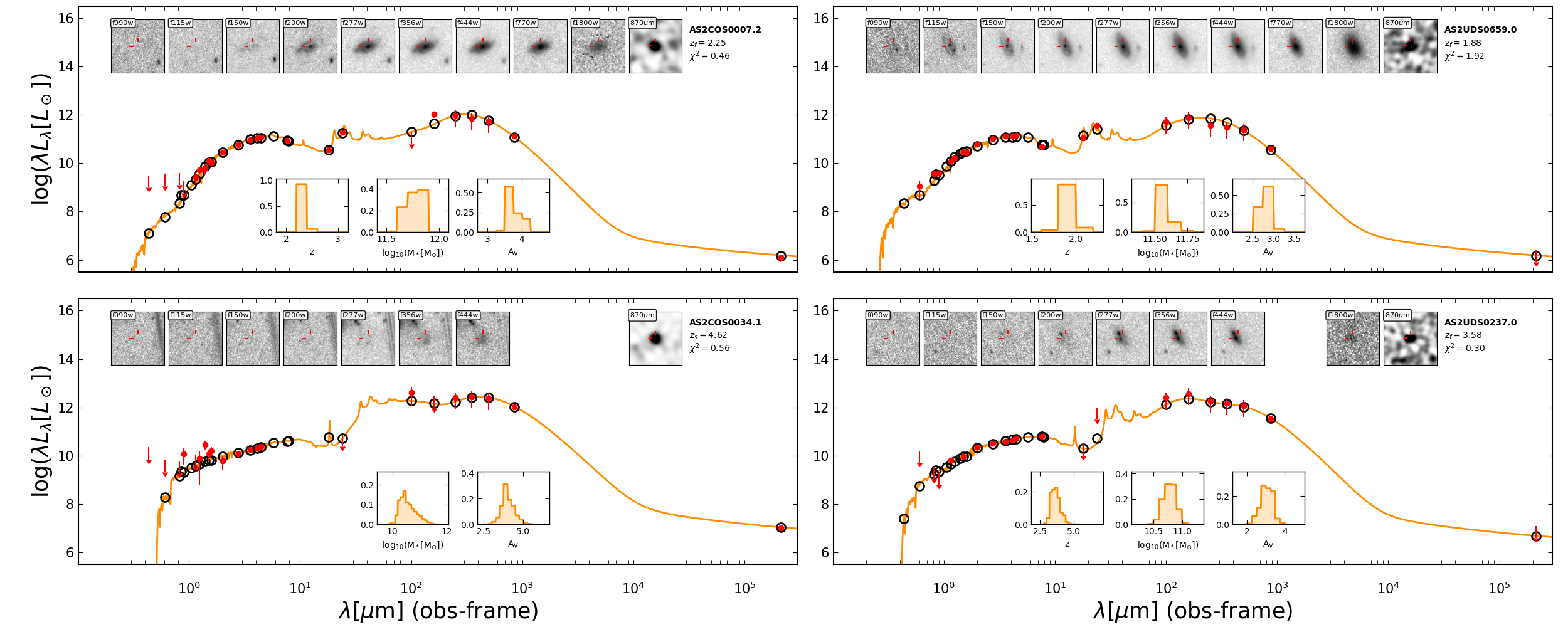}
    \caption{The multi-wavelength SEDs for four example SMGs that demonstrate the range of observed-frame infrared colours displayed by the SMG sample. Appendix B shows the SEDS for the full sample. For each source, we indicate the observed flux and error in each band with a red circle and errorbar.  Arrows show upper limits at 3$\sigma$.
    We over-plot the best-fit \texttt{\sc{magphys}} derived SED (orange line) and model fluxes (black circles). Below each SED we display the probability distributions of the {\sc{magphys}} derived properties: redshift (unless spectroscopic), stellar mass and A$_{ V}$, and report the best-fit redshift and reduced $\chi^2$ from the fit in the top right. Above the SED we display the multi-wavelength \jwst{} and ALMA 870\um\ imaging used as part of the analysis, with the SMG centroid indicated by the red tick markers.}
    \label{Fig:SED}
\end{figure*}

% Extraction
\subsection{Photometry}
\label{subsec:phot}
For \jwst\ and \hst\ bands we extract sources using {\textsc{sep}} \citep{Barbary2016}, a \textsc{python} version of \textsc{source extractor} \citep{Bertin1996}, with a noise-weighted combined long-wavelength (LW) stacked (F277W+F356W+F444W) NIRCam image as the detection image. For each source, we make a cutout in each of the \hst\ and \jwst\ bands, centred on this NIRCam long-wavelength detected source. Colour images made from the F444W/F356W/F277W NIRCam bands for the SMGs are shown in Figure \ref{Fig:Cimg}. Aperture photometry is performed in elliptical apertures with a  minimum diameter of $1\farcs0$ and corrected to the ``total'' values following \citet{Kron1980}. The aperture corrections are computed on the NIRCam LW stacked image and applied to all bands.  The MIRI observations were processed prior to the release of the updated photometric calibrations (August 2023). We thus scale the measured flux in the F770W and F1800W bands by 0.85 and 1.03 respectively following the \jwst\ documentation. We determine a median F444W AB magnitude for the SMG sample of $m_{\rm F444W}$\,=\,21.7\,$\pm$\,0.3 with a 16\qth\,--\,84\qth\ percentile range of  $m_{\rm F444W}$\,=\,20.3\,--\,23.1. 

%SPIRE, PACS, ALMA,VLA etc
We compile the mid-infrared to radio (5.8\um\ to 1.4GHz) photometry for our sources from \citet{Dud2020} and \citet{Simpson2020}. For the SMGs with no MIRI F770W and F1800W imaging, we adopt the IRAC channel 3 (5.8\um) and channel 4 (8\um) photometry, if IRAC channel 1 (3.6\um) flux is within 10 percent of the NIRCam F356W and IRAC channel 2 (4.5\um) flux is within 10 percent of the NIRCam F444W. We employ the IRAC photometry for two out of seven AS2COSMOS SMGs without MIRI coverage and 11 out of 38 AS2UDS SMGs.

\subsection{Final sample}

Having collated a sample of 88 AS2UDS and AS2COSMOS SMGs lying within the \jwst{} PRIMER fields, we visually inspect each of the galaxies' \jwst{} imaging to verify the association of the near-infrared counterpart to the ALMA source.  

We remove three UDS sources (AS2UDS0106.0, AS2UDS0175.0 and, AS2UDS0069.0) and three COSMOS sources (AS2COS0017.1, AS2COS0032.1 and, AS2COS0035.1) due to partial coverage in the \jwst{} NIRCam imaging. Of the remaining 82 SMGs, we identify one SMG, AS2UDS0490.0, with  no clear NIRCam counterpart, with a F444W magnitude of  $m_{\rm F444W}$\,=\,26.7\,$\pm$\,0.6 detected at 1.4$\sigma$. There is no MIRI coverage or any detection of this S$_{870}$\,=\,2.7\,mJy (S/N\,=\,4.5) source between UV and radio, thus we conclude it is likely spurious, with the ALMA catalogue of 707 sources from \citet{Stach2019} expected to have a two percent spurious fraction, and so exclude the source from our sample. We also remove two SMGs, AS2COS0005.1 and AS2COS0005.2, which appear strongly gravitational lensed and thus the derivation of their intrinsic properties including morphology would require detailed and uncertain lens modelling \citep[e.g.][]{Amvrosiadis2018,Bendo2023,Pearson2024}.
We also flag AS2UDS0281.0 and AS2UDS0382.0 as possibly weakly lensed, opting to keep them in the sample.

Finally, one ALMA source from \citet{Simpson2020}, AS2UDS00071.0, is deblended in the NIRCam F444W imaging, and high-resolution ALMA observations \citep{Gullberg2019}, into two sources which we label AS2UDS00071.0a and AS2UDS00071.0b. We model the morphology of AS2UDS0071.0a and AS2UDS0071.0b independently and scale the blended photometry by the flux ratio of the two sources. Thus in the final sample, we have 80 SMGs, for which the \hst{} and \jwst{} coverage for each SMG is detailed in Appendix A.

% magphys
\subsection{SED fitting}

Having defined a final sample of 80 SMGs with robust near-infrared counterparts, we use the SED fitting code Multi-wavelength Analysis of Galaxy Physical Properties and Photometric Redshift \citep[{\sc{magphys+photo-$z$}}, hereafter referred to as {\sc{magphys}};][]{daCunha2015, Battisti2019}, to derive the physical properties of each SMG\footnote{We note that we do not include an AGN component in the {\sc{magphys}} modelling as the fraction with significant AGN contributions is expected to be small \citep[e.g.,][]{Stach2019} (See Section \ref{Sec:Results})}. For 22 of the SMGs spectroscopic redshifts are available from the literature and ongoing millimetre and near-infrared spectroscopic surveys \citep[e.g.,][]{Mclure2018,Mitsuhashi2021}. For these sources we use the high-redshift version of {\sc{magphys}}, \citep[{\sc{magphys+high$z$ (v2)}};][]{daCunha2015, Battisti2019} with the redshift fixed to the spectroscopic redshift of each source, as detailed in Appendix A.

{\sc{magphys}} is a physically motivated SED fitting code that utilises the energy balance technique to fit the multi-wavelength photometry from UV to radio. This approach, as tested on SMGs by \citet{daCunha2015} and \citet{Dud2020}, and also on simulated galaxies \citep[e.g.,][]{Dud2020,Haskell2023}, models  the sub-millimetre and optical emission as originating from the same region of the galaxy. For a discussion of using the \texttt{\sc{magphys}} SED code to model high redshift SMGs, we refer the reader to \citet{Dud2020}. For consistencies with \citet{Dud2020}, for non-detections at wavelengths shorter than 8\um, we adopt a 3$\sigma$ upper limit while for those beyond 10\um, we use 1.5\,$\pm$\,1.0$\sigma$.

\begin{figure*}
    \includegraphics[width=\linewidth]{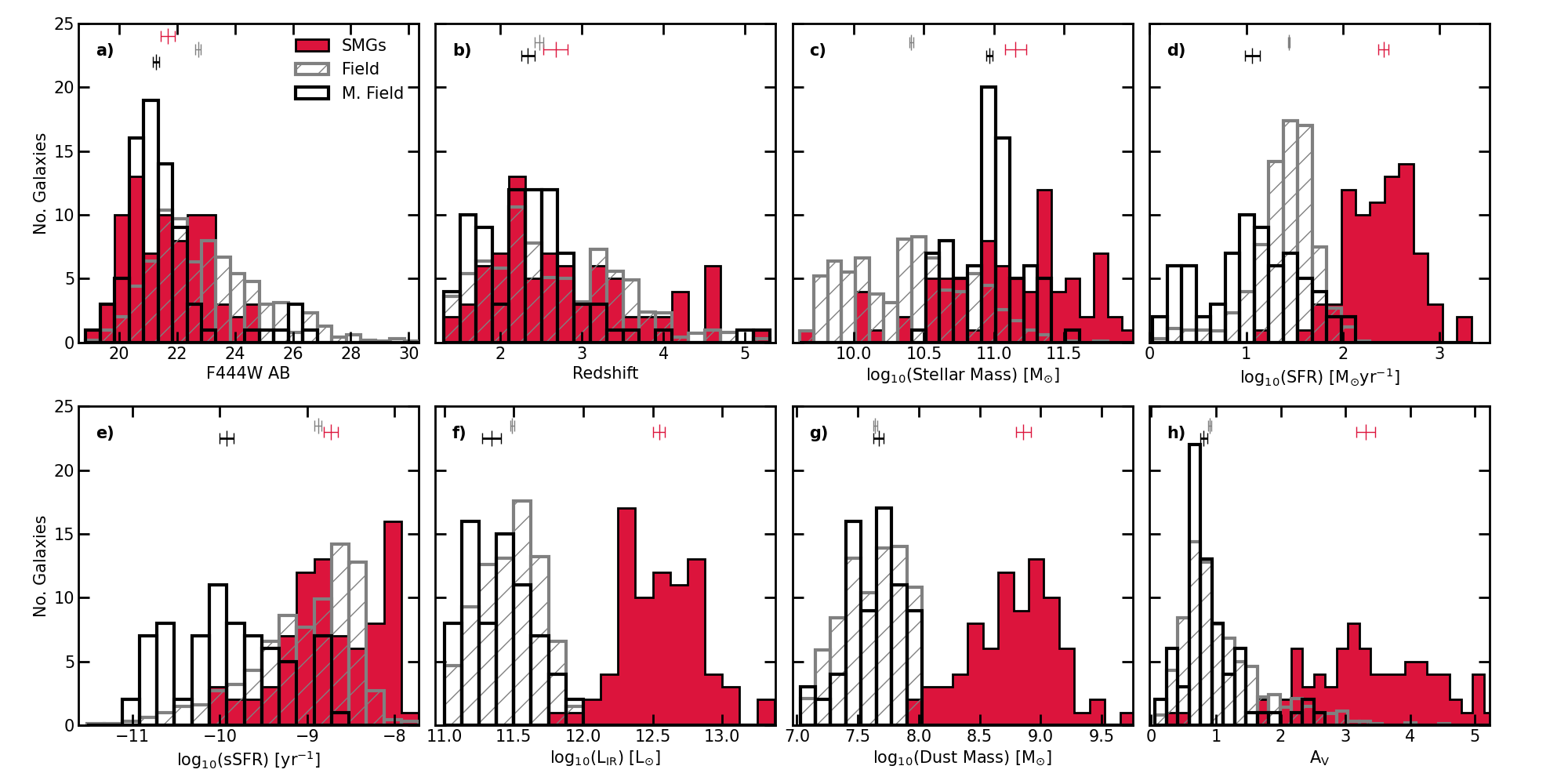}
    \caption{Distributions  of the \texttt{\sc{magphys}} derived SED properties for the SMG sample (\textit{red}) and \rev{field \textit{(hatched grey, scaled down by a factor 10)} and massive field (M. Field) samples \textit(open black histograms)}. We show the F444W AB magnitude \textit{(a)}, photometric redshift \textit{(b)}, stellar mass \textit{(c)}, star formation rate \textit{(d)}, specific star formation rate \textit{(e)}, infrared luminosity \textit{(f)}, dust mass \textit{(g)} and $V$-band attenuation (A$\rm_V$) \textit{(h)}. For each distribution, we indicate the bootstrapped median and uncertainty for SMGs (\textit{red marker and errorbar}) and the matched field  (\textit{grey marker and errorbar}) and massive field sample (\textit{black marker and errorbar}). By construction, the less massive field sample mirrors the SMG sample in redshift and specific star-formation rate distributions \rev{resulting in lower stellar masses, whilst the massive field sample more closely matches the SMGs distribution in stellar mass but with a lower median redshift. Both field samples have significantly lower dust masses, infrared luminosity, star-formation rates and \av.}}
    \label{Fig:SED_hist}
\end{figure*}

In Figure \ref{Fig:SED}, we show examples of the multi-wavelength photometry and \texttt{\sc{magphys}} SED fits, with the fits for the whole sample presented in Appendix B. We summarise some of the key derived physical properties of the SMG sample in Figure \ref{Fig:SED_hist}. We estimate a median redshift of $z$\,=\,2.70\,$\pm$\,0.15 with a (16\qth\ to 84\qth\ percentile range of $z$\,=\,1.9\,--\,3.9), which is comparable to that derived by \citet{Dud2020} for AS2UDS of $z$\,=\,2.6\,$\pm$\,0.8 and the median redshift of the  AS2COSMOS survey of $z$\,=\,2.7\,$\pm$\,0.9 as derived by \citet{Simpson2020}, as well as other studies of SMGs at similar 870\um\ flux densities\citep[e.g.,][]{daCunha2015,Ling2022}. From the {\sc{magphys}} SED fitting, we derive a median stellar mass for the SMGs of $\rm \log_{10}(M_\ast[M_{\odot}])$\,=\,11.20\,$\pm$\,0.10 with a 16\qth\,--\,84\qth\ percentile range of $\rm \log_{10}(M_\ast[M_{\odot}])$\,=\,10.6\,--\,11.6, and a median specific star-formation rate of  $\rm \log_{10}(sSFR[yr^{-1}])$\,=\,$-$8.7\,$\pm$\,0.1 with a 16\qth\,--\,84\qth\ percentile range of $\rm \log_{10}(sSFR[yr^{-1}])$\,=\,$-$9.2 to $-$8.0. We derive a median dust mass of $\rm \log_{10}(M_d[M_{\odot}])$\,=\,8.8\,$\pm$\,0.1 with a 16\qth\,--\,84\qth\ percentile range of $\rm \log_{10}(M_d[M_{\odot}])$\,=\,8.4\,--\,9.1 and a median infrared luminosity of $\rm \log_{10}(L_{IR}[L_{\odot}])$\,=\,12.6\,$\pm$\,0.1 with a 16\qth\,--\,84\qth\ percentile range of $\rm \log_{10}(L_{IR}[L_{\odot}])$\,=\,12.3\,--\,12.8. These are consistent with the sample of 707 SMGs in AS2UDS, for which \citet{Dud2020} derived a median stellar and dust mass of $\rm \log_{10}(M_\ast[M_{\odot}])$\,=\,$11.1^{+0.2}_{-0.3}$ and  $\rm \log_{10}(M_d[M_{\odot}])$\,=\,$8.8^{+0.4}_{-0.4}$ and a median specific star-formation rate and infrared luminosity of  $\rm \log_{10}(sSFR[yr^{-1}])$\,=\,$-$8.73\,$\pm$\,0.04 and  $\rm \log_{10}(L_{IR}[L_{\odot}])$\,=\,12.45\,$\pm$\,0.02

As expected given that the \jwst{} photometry agrees with earlier IRAC and \hst\ observations, while typically being more sensitive, the basic derived properties of the SMGs don't alter significantly when \jwst\ photometry is included. The main benefit of deeper observations at 1\,--\,3\um\ is improved constraints on the presence of Balmer breaks at $z$\,$\sim$\,1\,--\,6.

\subsection{Field  sample}
To investigate the physical mechanism that drives and differentiates our SMG sample from the typical star-forming population at the same epochs ($z$\,=\,2\,--\,5), we compare their multi-wavelength properties to two samples of less active field galaxies. We construct the field samples from the $K$-band selected sample analysed by \cite{Dud2020} in the UKIDSS UDS field  for which deep 22-band photometry is available covering the UV/optical through to the far-infrared/submillimeter and radio. \cite{Dud2020} used the {\sc magphys} code to derive photometric redshifts, stellar masses and other physical properties for the $\sim$\,300,000 $K_{\rm AB}$\,$\leq$\,25.7 galaxies in this $\sim$\,0.8\,degree$^2$ field  from the UKIDSS survey (Almaini in prep.). 

Starting from this field catalogue we first select galaxies with no photometric flags that lie within the footprint of the \jwst\ PRIMER survey and have estimated dust masses and far-infrared luminosities that ensure they are less actively star-forming than our SMG sample:  $M_{\rm dust}$\,$=$\,10$^7$—10$^8$\,M$_\odot$ and  $L_{\rm IR}$\,$=$\,10$^{11}$—10$^{12}$\,L$_\odot$.  We then bin the redshift distribution of the SMGs and field  galaxies in  $\Delta z$\,$=$\,0.5 bins and in each bin we rank the field  galaxies based on their stellar masses.  We then determine the number of SMGs for each bin and select ten times as many field  galaxies from the corresponding redshift bin, starting with the most massive and going down the stellar-mass ranked list. In this way, we construct a field control sample that accurately matches the redshift distribution of our SMG sample and comprises the most massive galaxies at each epoch that have dust masses and far-infrared luminosities below those determined for the SMG population. We adopt the  photometry and \texttt{\sc{magphys}} SED fitting results from \citet{Dud2020}. 

In Figure \ref{Fig:SED_hist}, we show the distribution of physical properties for the UDS field  sample in comparison to the SMGs. By construction, the field  sample has similar redshift and specific star-formation rates to the SMG sample, whilst having significantly lower dust masses and A$_{\rm V}$. The field  sample has a median redshift and median absolute deviation  of $z$\,=\,2.5\,$\pm$\,0.1, whilst their specific star-formation rates are sSFR\,=\,$-$8.88\,$\pm$\,0.04 $\rm yr^{-1}$ and  \av\,=\,0.90\,$\pm$\,0.03. The field  sample is also fainter in F444W magnitude with a median value of $m_{\rm F444W}^{\rm field}$\,=\,22.74\,$\pm$\,0.10 compared to $m_{\rm F444W}^{\rm SMGs}$\,=\,21.70\,$\pm$\,0.30, which \rev{is reflected in the lower median stellar mass of $\rm \log_{10}(M_\ast[M_{\odot}])$\,=\,10.4$\pm$\,0.1. Unfortunately, it is not possible to select a large sample of less active star-forming galaxies that are both matched to the SMG's stellar mass and redshift distributions, because most massive galaxies at $z$\,$\geq$\,2\,--\,3 are also SMGs.}

\rev{To address the mass offset of the field sample we define a second ``massive’’ field sample. We again select galaxies with no photometric flags lying within the PRIMER survey, but now require that they have estimated dust masses below $M_{\rm dust}$\,$\leq$\,$10^8$\,M$_{\odot}$ (to ensure they are distinct from the SMGs) and stellar masses of $M_\ast$\,$\geq$\,$3.2\times10^{10}$\,M$_{\odot}$ (to more closely match the stellar masses of the SMGs).  Finally, the binned redshift distribution was used to attempt to maximise both the size of the sample and its median redshifts, while also having a comparable breadth to the SMG redshift distribution.  This resulted in a final ``massive'' field sample comprising 80 galaxies with a median stellar mass of $\log_{10}(M_\ast[M_\odot])$\,$=$\,11.0\,$\pm$\,0.1, a median far-infrared luminosity of $\log_{10}(L_{\rm IR}[L_\odot])$\,$=$\,11.3\,$\pm$\,0.1 and a median redshift of $z$\,$=$\,2.3\,$\pm$ 0.1 (16\,--\,84$^{\rm th}$ percentile range of $z$\,$=$\,1.7\,--\,2.9).  Hence this second field sample is comparable in mass to the SMGs and comprises much less actively star-forming galaxies, but lies at somewhat lower redshifts than the SMGs.  The {\sc magphys} derived SED properties of the SMGs, field and ``massive’’ field samples are shown in Figure \ref{Fig:SED_hist}.}

\subsection{Morphological analysis}
%morphology
In Figure \ref{Fig:Cimg} we show NIRCam F277W/F356W/F444W colour images of the SMGs (ordered by decreasing 870\um\ flux), highlighting the diverse range of rest-frame optical -- near-infrared morphologies from faint and red galaxies, interacting and merging systems to disks and regular  spirals. These near-infrared observations sample the galaxy population at $\leq$1\um\ in the rest frame at $\approx$\,1\,kpc resolution and thus provide insights into the structure and morphology of the stellar continuum emission which are much less affected by dust than previously possible. 

We visually assess as well as quantify the rest-frame near-infrared morphology of the SMGs and field  galaxies using non-parametric and parametric analyses of the \jwst\ observations. We exclude the \hst\ observations from our morphological analysis because at the median redshift of our sample ($z$\,$\sim$2.7), \hst\ observations only sample the rest-frame UV/optical emission of the galaxies where SMGs are inherently faint. The longer wavelength \hst\ observations sampling the observed frame $\simeq$1.6\um\ emission is also well covered by higher signal to noise \jwst\ observations. 

%cutouts
The parametric and non-parametric analysis of the \jwst\ imaging employs 4\farcs{0}\,$\times$\,4\farcs{0} cutouts ($\simeq$ 30\,kpc square) of each source with a pixel scale of 0\farcs{0}4 per pixel. The cutout in each filter is centred on the source detection in the NIRCam long-wavelength (F277W+F356W+F444W) stacked image, as detailed in Section \ref{subsec:phot}. We first smooth the source segmentation map generated from \texttt{\sc{sep}} (see Section \ref{subsec:phot}) using the binary dilation routine in {\sc{photutils}} \citep[][]{Phot2022}. Then, excluding this segmented region, we mask the remaining sources in the cutout, down to a 1$\sigma$ isophote. This ensures full masking of any contaminants (spurious or otherwise)\footnote{We note the simultaneous modelling of nearby sources returns similar parametric results.}. We further use {\sc{photutils}} to model (and remove) the background level in each cutout as well as to quantify the root-mean-square (rms) noise. In the following sections ``cutout'' refers to these 4\farcs{0}\,$\times$\,4\farcs{0}, background subtracted and  masked image that is used in the morphological analysis that follows.

%PSF
Prior to measuring the morphology of the galaxies we first derive the point spread function (PSF) for each of the \jwst\ bands.  We use {\sc{webbpsf}} (version 1.2.1; \citealt{Perrin2014}) to generate PSF models for the MIRI and NIRCam detectors (for both short- and long-wavelength channels) which are calibrated with wavefront models at the epoch of the observations. The PSFs for each NIRCam and MIRI filter are then inserted into individual exposures of the final mosaic and drizzled to the final world coordinate system  solution, producing a field  of view averaged PSF model.

\subsubsection{Visual morphology}
\begin{figure*}
\begin{minipage}[c]{0.5\linewidth}
\includegraphics[width=\linewidth,trim={0cm 0cm 15cm 0cm},clip]{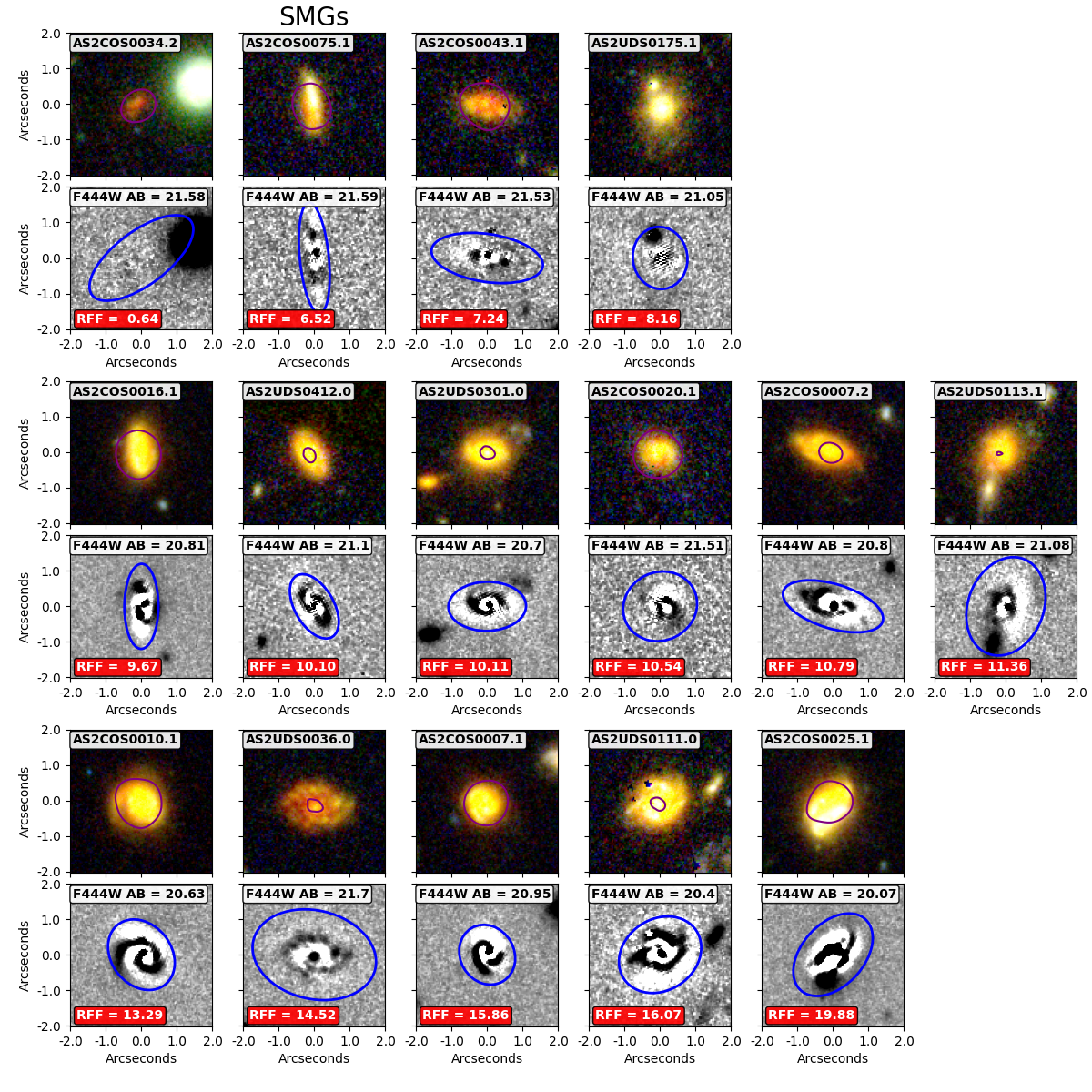}
\end{minipage}
\hfill
\begin{minipage}[c]{0.5\linewidth}
\includegraphics[width=\linewidth,trim={0cm 0cm 15cm 0cm},clip]{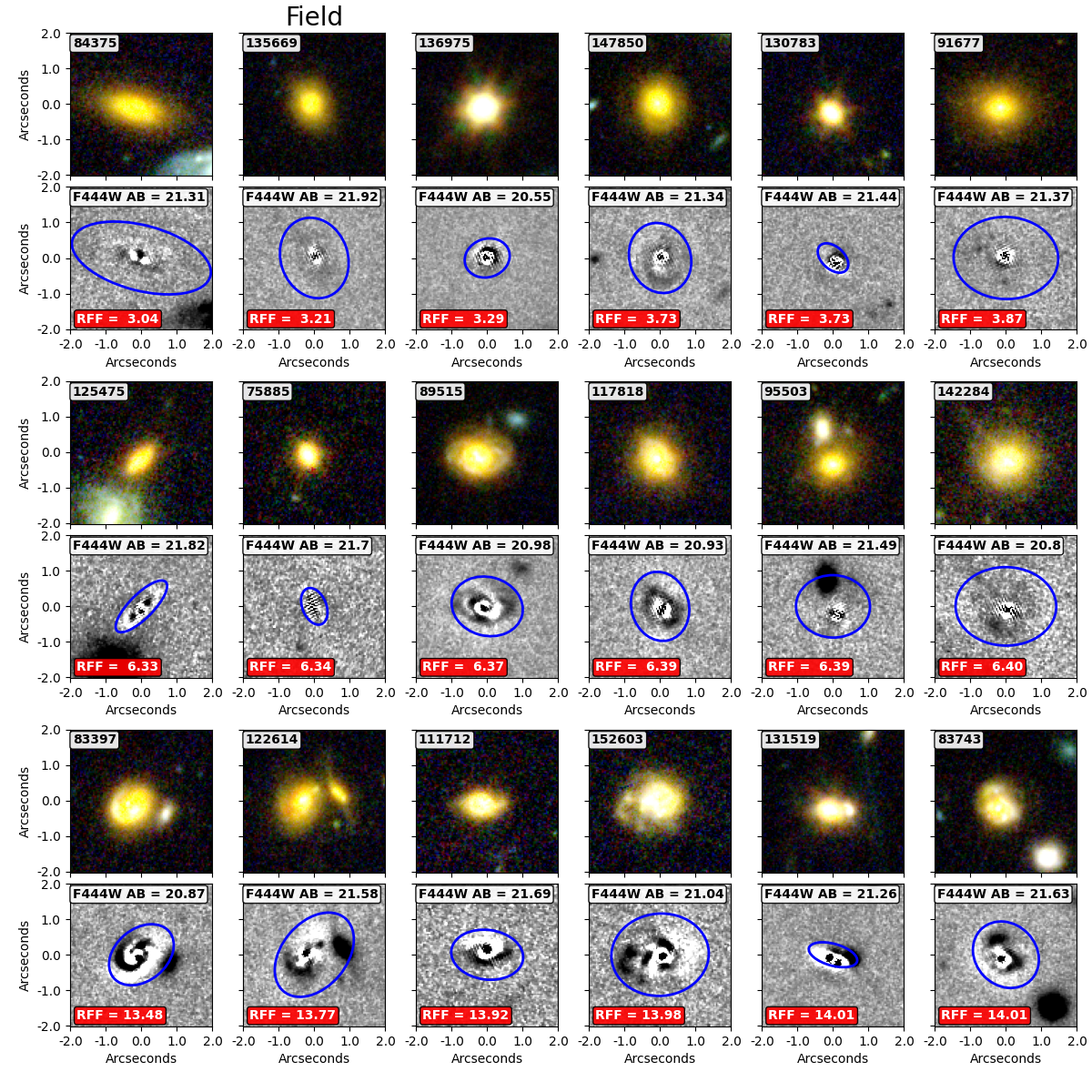}
\end{minipage}%
\caption{For both SMGs (\textit{left}) and field  galaxies (\textit{right}) we show the examples of the F356W/F277W/F115W RGB-colour image. For the SMGs we overlay the ALMA 870\um\ 11$\sigma$ contour (\textit{purple}) from \citet{Stach2019} or \citet{Simpson2020}. In the lower panel of each source, we show the residuals of the  {\sc{galfitm}} single S\'ersic model fit. The blue ellipse marks the region within which we calculate the residual flux fraction (RFF) of the S\'ersic fit, as quantified in the lower-left corner of each panel. The galaxies are selected to have a $m_{\rm F444W}$\,=\,20\,--\,22 and are ranked from low-to-high RFF. No clear distinction is identified in the F444W RFF values between the SMGs and field  galaxies.}
\label{Fig:RFF}
\end{figure*}

We first undertake a crude initial visual assessment of the morphologies of the SMGs and field samples.
This involved inspection of both the colour images and {\sc{galfit}} residual maps in the F444W band (Figure \ref{Fig:Cimg} \& \ref{Fig:RFF}) of the ALMA counterpart to identify the overall structure and any distorted morphologies, asymmetric structures or potential tidal features.  In addition, we assessed the presence of companions (either within $\sim$1\,mag in brightness of the target galaxy, or fainter) within a wider 10\farcs{0}\,$\times$\,10\farcs{0} region (out to a radius of $\simeq$40\,kpc at $z\sim2.7$).  Strongly disturbed galaxies, those with tidal features or disturbed galaxies with bright companions were classed as “major” mergers (M), while less disturbed galaxies or those with asymmetries and fainter companions  were  classed as potential “minor” mergers \rev{or interacting (I) in addition to undisturbed disks (D) and compact (C) sources. These classifications are marked in Figure \ref{Fig:Cimg}.}

\subsubsection{Non-parametric morphology}

%non-parametric
To measure the half-light radii of the SMGs, we employ a curve of growth approach that makes no assumptions about the underlying structure of the galaxy's light distribution. We adopt this approach because SMGs have long been assumed to originate from merger-driven events with clumpy unstable gas-rich disks, \citep[e.g.,][]{Smail1998,Greve2005,Tacconi2008,Engel2010} and therefore the morphologies of some sources may deviate from simple parametric profiles.

%identifying b/a and PA
We first perform a curve of growth analysis in each of the \jwst{} bands from 0.9\um\ to 18\um. This is achieved by fitting a Gaussian profile to the cutout of each galaxy to determine its image parameters, allowing the centroid ($x,y$), axis ratio ($b/a$) and position angle (PA) to vary. We note the original centroid of the cutout is derived from the \texttt{\sc{sep}} source detection on the stacked F277W+F356W+F444W NIRCam bands (Section \ref{subsec:phot}), and thus may not be the apparent centre of the galaxy at shorter wavelengths. A curve of growth is then derived in each band using ellipses which align to the galaxy's axis ratio and position angle. From the curve of growth, we measure the convolved radii containing 20, 50 and 80 per cent of the flux of each galaxy. The intrinsic radii of the galaxies are derived by de-convolving the sizes with the PSF in each band, measured through a similar curve of growth analysis.

%statmorph
To provide more quantitative, non-parametric, morphological indicators, we use the \texttt{\sc{statmorph}}\footnote{\url{https://statmorph.readthedocs.io/en/latest/}} code \citep{statmorph2019} which measures the Concentration, Asymmetry and Clumpiness ($C$, $A$, $S$; \citealt{Abraham2003,Lotz2008}) parameters that quantify how concentrated, asymmetrical and clumpy the galaxies' surface brightness profiles are, with higher values indicating more concentrated, asymmetric, or clumpier light profiles. We run {\sc{statmorph}} on both SMG and field  samples in all \jwst{} bands using the same segmentation maps and PSFs as for the growth curve analysis described above. 

In addition, the  Gini and M$_{20}$ parameters are also derived (for full definitions see \citealt{Lotz2004,Synder2015}). Briefly, the Gini parameter defines the pixel distribution of the galaxy's light, where $G$\,=\,1 corresponds to all of the light  concentrated in one pixel whilst $G$\,=\,0 indicates each pixel contributes equally. The M$_{20}$ parameter measures the second moment of the brightest 20 percent of pixels in the galaxy. This is normalised by the total  moment for all pixels.  Highly negative values indicate a high concentration of light, not necessarily at the centre of the galaxy. To validate the robustness of the \texttt{\sc{statmorph}} measurements we compare the half-light radius to that derived from our curve of growth analysis.  We derive a median curve of growth to \texttt{\sc{statmorph}}  median half-light radius ratio of $R_{\rm h,CoG}$\,/\,$R_{\rm h,statmorph}$\,=\,1.06\,$\pm$\,0.01 with a 16\qth\,--\,84\qth\ percentile range of $R_{\rm h,CoG}$\,/\,$R_{\rm h,statmorph}$\,=\,0.93\,--\,1.45 for the field  sample in the F444W band. This indicates good agreement between the two independent morphological measurements\footnote{We note however we expect some variation between the two methods due to the definitions of centroid and total fluxes used (see \citealt{Lotz2004})}. 

\subsubsection{Parametric morphology}

To quantify both the parametric morphological profiles of the galaxies  and the deviations from these, we use the \texttt{\sc{galfitm}} code \citep{Haubler2013}. \texttt{\sc{galfitm}} is a Python-wrapper for \texttt{\sc{galfit}} \citep{Peng2010}, that allows multi-component parametric models to be fit to a galaxy's multi-wavelength light distribution. For our analysis we use a single S\'ersic model, convolved with the PSF of the relevant \jwst\ band. We fit each band independently thus allowing us to constrain the intrinsic wavelength dependence of the galaxy's morphology. 

To constrain the accuracy of the parametric analysis, we compare the {\sc{galfitm}} S\'ersic index ($n$) to that derived by \texttt{\sc{statmorph}} for the field  galaxies in the NIRCam F444W band. We establish a median ratio  of $n_{\rm F444W}^{\rm GalfitM}$\,/\,$n_{\rm F444W}^{\rm statmorph}$\,=\,1.00\,$\pm$\,0.01 and a 16\qth\,--\,84\qth\ percentile range of  $n_{\rm F444W}^{\rm GalfitM}$\,/\,$n_{\rm F444W}^{\rm statmorph}$\,=\,0.81\,--\,1.26, indicating good agreement between the two independent codes. 

Although a parametric S\'ersic fit can model the overall distribution of a galaxy's light profile, it is the deviations (residuals) from this simple profile, as shown in Figure \ref{Fig:RFF}, that encode important information about the detailed structure of the galaxies. To quantify the residuals we use the 
residual flux fraction (RFF), as defined in \citet{Hoyos2011,Hoyos2012},
\begin{equation}
    \rm RFF\,=\,\frac{\sum_{i,j \in A}|I_{i,j}\,-\,I_{i,j}^{model}|\,-\,0.8\,\times\,\sum_{i,j \in A}\,\sigma_{Bkg\,i,j}}{\sum_{i,j \in A} I_{i,j}}
\end{equation}
where the sum is performed over all pixels within 2.5 times the Kron radius, as derived in Section \ref{subsec:phot}. $\rm |I_{i,j}\,-\,I_{i,j}^{model}|$ is the absolute value of pixel $i,j$'s residuals to model S\'ersic model fit, whilst $\sum_{i,j \in A}I_{i,j}$ indicates the total flux measured in the source as defined in Section \ref{subsec:phot}. The 0.8 factor multiplied by the sum over the background rms of the region ($\sigma_{Bkg\,i,j}$), ensures that a blank image with constant variance has a RFF\,=\,0.0 (for details see \citealt{Hoyos2012}). 

\begin{figure*}
 \includegraphics[width=\linewidth]{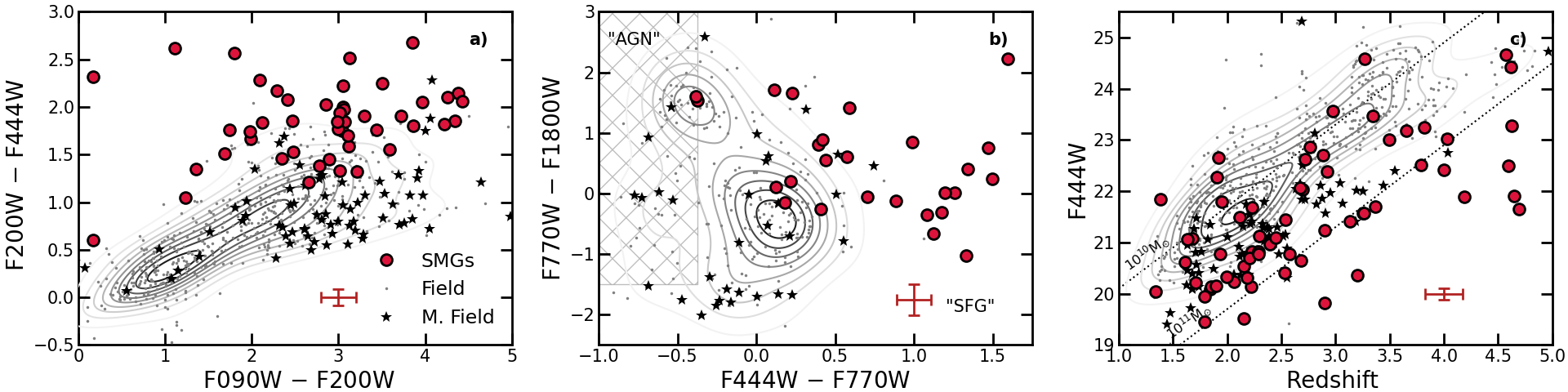}
    \caption{The distribution of \textit{(a)} NIRCam colours, \textit{(b)} MIRI colours and \textit{(c)} F444W AB magnitude as a function of redshift , with lines of constant stellar mass overlaid in the latter. A representative error bar is indicated in the lower-right corner of each panel. Panel \textit{a)} demonstrates the SMGs are much redder than \rev{our massive and sSFR-matched field populations}, as expected. Panel \textit{b)} highlights that a fraction of SMGs and field galaxies may host AGN with a very red F770W$-$F1800W colour and bluer F444W$-$F770W indicating a strong upturn in the SED beyond rest-frame 3\um. We label the ``AGN" and star-forming galaxy (``SFG") regions in the colour space adapted from \citet{Kirkpatrick2013,Kirkpatrick2017}. We note the F770W$-$F1800W\,$>$\,1 colour can also be driven by the 6.2\um\ PAH feature detected in the F18000W filter at $z$\,$\sim$\,1.9. Panel \textit{c)} demonstrates that the SMGs are the most massive (and so typically also brightest) galaxies at their epoch \rev{where the two field samples have been selected to be the most massive low far-infrared luminosity galaxies.}}
    \label{Fig:Colours}
\end{figure*}

%uncertanities
It is well known that morphological codes such as \texttt{\sc{galfitm}} and \texttt{\sc{statmorph}} systemically underestimate the uncertainties on the derived morphological parameters \citep[e.g.,][]{VW2012,VW2024}. To overcome this we employ an empirical approach which utilises the unique wavelength coverage of the NIRCam observations in the PRIMER Survey. Specifically, the majority of field galaxies (816/850) and SMGs (75/80) in our sample have both F410M and F444W observations. The F410M is a medium band filter covering the observed-frame 3.8\,--\,4.3\um\ emission, whilst the wide-band F444W filter is sensitive to the 3.8\,--\,4.9\um\ emission. Thus, for each galaxy with observations in both filters, we can make two independent measures of the galaxy's morphology at very similar rest-frame wavelengths. Analysing the variance in morphological parameters between the two filters as a function of signal-to-noise allows us to infer a representative uncertainty on the morphological parameters at a given wavelength for each source given its signal-to-noise in that band. We use this conservative approach to quantify the uncertainties in the following analysis.

\section{Results} \label{Sec:Results}

From an initial sample of 88 SMGs, we have constructed a sample of 80 not-strongly-lensed SMGs with reliable (S/N\,$>$\,4.5) ALMA 870\um\ detections and multi-wavelength \hst\ and \jwst\ coverage from 0.4\,--\,18\,\um. Full details of the wavelength coverage for the individual SMGs in our sample are given in Appendix A. From the initial sample, we omitted two strongly lensed SMGs due to uncertainties on the photometry and structural properties introduced by the lensing configuration of each source, along with flagging two candidate weakly lensed sources, AS2UDS0281.0 and AS2UDS0382.0 (Figure \ref{Fig:Cimg}). Identifying 5\,$\pm$\,3$\%$ of ALMA-detected SMGs in our sample are affected by foreground galaxy lensing matches the prediction of \citet{Chapman2002}, that between 3 and 5 per cent of sub-millimetre sources identified in blank-field SCUBA observations would be affected by lensing.

For the 60 SMGs with \hst\ coverage, we detect no F160W counterpart at S/N\,$>$\,3 in 10 sources (17\%). All of these have clear near-infrared counterparts in the NIRCam F444W band at S/N\,$>$\,5, whilst three also have a S/N\,$>$\,3 in the NIRCam F150W band. This leaves seven ``NIR-faint'' (sometimes imprecisely described as “\hst -dark”) SMGs with no detection in the $H$-band (S/N\,$<$\,3). These seven SMGs have significant levels of dust attenuation as derived from \textsc{magphys} with \av\,$>$\,3.9. This indicates that whilst the significant dust content of these galaxies can lead to them being undetected in \hst\ observations, it is also driven by the lack of sensitivity and depth of previous observations. 

Of the 78 SMGs with NIRCam short-wavelength (SW) coverage, we identify one galaxy, AS2UDS0346.0 at $z$\,=\,4.1, that is undetected (S/N\,$<$\,3) blueward of 2.7\um\, and is first detected by \jwst\ in the F277W band at S/N\,$\geq$\,5. From our {\sc{magphys}} analysis we derive significant dust attenuation with \av\,=\,5.1$^{+2.0}_{-0.6}$ for AS2UDS0346.0 with a stellar mass of $\log_{10}(M_\ast[M_{\odot}])$\,=\,11.2$^{+0.1}_{-0.2}$ that is representative of our sample. For SMGs with suitable NIRCam coverage we identify detections (S/N\,$>$\,3) in F200W for 96\% (70/73), in F150W for 87\% (68/78) and in F090W for 59\% (43/73) of the sources respectively. All have  S/N\,$>$\,3 detections in the NIRCam-LW and MIRI bands. 

In Figure \ref{Fig:Colours} we investigate the photometric properties of the SMGs in the \jwst\ bands by comparing the NIRCam and MIRI colours. Specifically, in Figure \ref{Fig:Colours}a we compare the NIRCam F200W$-$F444W colour with the F090W$-$F200W colour. At the median redshift of the SMG and field  galaxies, these filters sample the rest-frame $UVJ$ regime of the galaxies' spectral energy distributions, highlighting the distinct red nature of the SMGs compared to the general field. Given this, it may be possible by colour selection to identify the \jwst\ counterpart to the sub-millimetre bright SCUBA-2 source without requiring higher resolution ALMA observations \citep[e.g.,][]{Stacey2013, Chen2016, An2018,Hwang2021}. To investigate this, for each SMG we compare the NIRCam F200W\,--\,F444W colour of all sources detected in the 14\farcs{8} diameter SCUBA-2 beam to that of the ALMA-selected SMG. However only in 51\% (41/80) of the SCUBA-2 observations, the ALMA source represents the reddest source in the beam. This indicates that whilst colour is a good identifier of SMGs, further information such as proximity to source centroid, predicted 870\um\ fluxes and multi-wavelength observations from infrared to radio are required to isolate the exact near-infrared counterpart \citep[e.g.,][]{Downes1986,Ivison2002,Hodge2013,Casey2013,An2018,Gillman2023}.

Figure \ref{Fig:Colours}b compares the MIRI and NIRCam colours with the F770W$-$F1800W colour shown as a function of F444W$-$F770W colour. At the median redshift of the SMG and field  galaxies, these filters sample the rest-frame near and mid-infrared. The SMGs on average have redder F444W$-$F770W colours with comparable F770W$-$F1800W colours to the field population, reflecting the dust attenuation of the shorter wavelength light. The SMGs and field  galaxies which exhibit very red F770W$-$F1800W colours with bluer F444W$-$F770W colour may indicate the presence of Active Galactic Nuclei (AGN) as noted by \citet{Ivison2004} who used MIPS/24\um\ and IRAC colours to identify AGN and starburst galaxies. We highlight the regions of the colour diagram that an ``AGN'' or star-forming galaxy (``SFG'') would likely inhabit, adapted from \citet{Kirkpatrick2013,Kirkpatrick2017}. Of those with MIRI coverage, we identify 24\,$\pm$\,5\% (69/282) of the field galaxies \rev{(and 21\,$\pm$\,3\% (7/33) of the massive field sample)} have near-infrared colours that may indicate the presence of AGN activity whilst for the SMGs we derive a lower fraction of 3\,$\pm$\,1\% (2/37). The two SMGs with ``AGN'' like colours, AS2UDS0259.0 and AS2UDS0659.0, both exhibit extended morphologies in the MIRI F770W and F1800W observations with no visible point source, indicating any contribution from an AGN is minimal. This is in agreement with previous studies which identify the AGN contribution in SMGs to be small \citep[e.g.,][]{Stach2019}. There are four SMGs with a compact, point-source like morphology. AS2UDS0173.0 which is not covered by the MIRI observations as well as AS2UDS287.1, AS2UDS509.0 and AS2UDS516.0. All sources appear point-source like in all \jwst\ bands, suggesting they  may just be compact galaxies as indicated in Figure \ref{Fig:Cimg}. The SFG and AGN classification in Figure \ref{Fig:Colours}b does not fully encapsulate the bi-modality present in the colour space, especially in the field galaxies. Galaxies with F1800W$-$F770W$>$1 and F444W$-$F770W$<$0, represent a bright ($m_{\rm F444W}$\,=\,21.5\,$\pm$\,0.1) lower redshift subset ($z$\,=\,1.9\,$\pm$\,0.04) with significantly brighter F1800W emission ($m_{\rm F1800W}$\,=\,20.3\,$\pm$\,0.1) compared to the ``bluer'' population ($z$\,=\,2.9\,$\pm$\,0.1, $m_{\rm F444W}$\,=\,22.9\,$\pm$\,0.1, $m_{\rm F1800W}$\,=\,23.3\,$\pm$\,0.1). At $z$\,=\,1.9, the F1800W filter is sensitive to the  6.2\um\ PAH feature, which can strongly enhance the mid-infrared emission in massive galaxies, resulting in red F770W$-$F1800W colours \citep[e.g.,][]{Draine2007,Aniano2020, Shivaei2024}.

In Figure \ref{Fig:Colours}c we show the correlation between NIRCam F444W magnitude and the redshift of the galaxies. We overlay lines of constant stellar mass, highlighting that the SMGs are the most massive galaxies at their epoch \citep{Dud2020}, \rev{the massive field sample being comparable and the less massive field sample exhibiting lower stellar masses at all epochs.}

\subsection{Near-infrared morphology}

While SMGs have long been known to exhibit complex, potentially merger-like, morphologies based on rest frame $UV$ imaging \citep[e.g.,][]{Smail1998,Chapman2004,Swinbank2010,Aguirre2013} the influence of dust has meant their true stellar morphologies are still unknown. The combination of near-infrared colour images (Figure \ref{Fig:Cimg}) and F444W S\'ersic fit residual maps (Figure \ref{Fig:RFF}) provides an unprecedented insight into the stellar structures and asymmetries in the SMGs. 

From a visual inspection, in 21/80 SMGs,, we identify clear spiral arm features in the S\'ersic model residuals, as well as clear clumpy structures in 30\,$\pm$\,4\% (24/80) of the SMGs. In particular, one galaxy AS2UDS0259.0, shows a bar-like structure in the F444W image indicating complex stellar structures may be present, similar to those identified in recent \jwst\ and ALMA studies \citep[e.g.,][]{Hodge2019,Smail2023,Rujopakarn2023,Zhaoxuan2023}.

We estimate that 40\,$\pm$\,10\% (32/80) of the SMGs are isolated systems, with no strong perturbations or obvious bright neighbours on $\simeq$20\,--\,30\,kpc (projected) scales, while 16 SMGs (20\,$\pm$\,5\%) in our sample have potential companions or are interacting with another massive galaxy suggesting major mergers, the remaining 32 SMGs (40\,$\pm$\,10\%) have faint companions and show signs of disturbance (e.g., asymmetries in the  {\sc galfit} F444W residuals) that suggest potential minor interactions and mergers. There are at least two such examples in our sample of SCUBA-2 sources, S2UDS0322 and S2COS0034, and one ALMA source (AS2UDS0071.0) that comprise multiple ALMA-detected, or NIRCam-F444W detected, galaxies that show clear signs of galaxy interactions with S2COS0034 comprising two SMGs that are both detected in [C{\sc ii}] emission at $z_{\rm [CII]}$\,=\,4.62 \citep{Mitsuhashi2021}.

The median far-infrared luminosity of the candidate major mergers is $L_{\rm IR}$\,$=$\,10$^{12.53\pm0.05}$\,$L_\odot$, compared to $L_{\rm IR}$\,$=$\,10$^{12.49\pm0.05}$\,$L_\odot$ for the remainder of the sample, indicating no strong dependence of far-infrared luminosity on merger state \citep[e.g.,][]{Hopkins2010}.  Indeed, an equivalent visual classification of the less active field sample returns very similar merger statistics with 25\,$\pm$\,5\% of the sample showing evidence for potential major mergers and the remaining $\sim$\,75\% showing much weaker or no evidence for merger-related disturbance or companions on $\simeq$\,20--30\,kpc scales. This similarity in the rate of mergers between SMGs and less active galaxies is consistent with the theoretical results from the {\sc eagle} simulation from \cite{Mcalpine2019} who found that simulated SMGs had similar rates of recent mergers to less active galaxies, in part because most galaxies at high redshifts are undergoing continuous infall and merging.

\begin{figure*}
    \includegraphics[width=\linewidth]{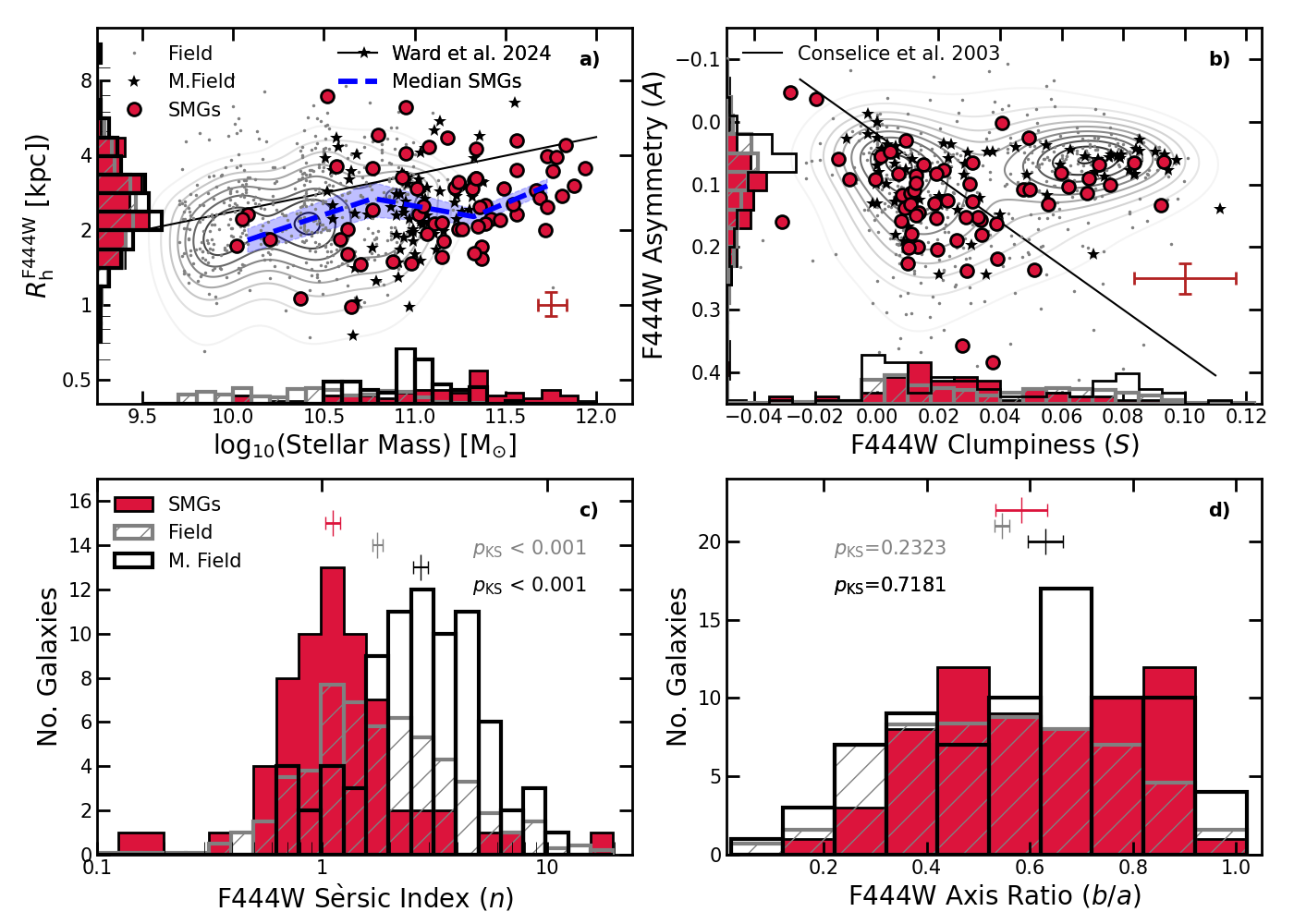}
    \caption{The correlation between \textit{a)} stellar mass and half-light radius, and \textit{b)} the Clumpiness ($S$) and Asymmetry ($A$) for the SMG and field samples in the NIRCam F444W filter. In each panel we show histograms of each parameter for the \rev{field (\textit{hatched grey histogram, scaled down by a factor 10}), massive field (\textit{black})} and SMG sample (\textit{red}). In panel \textit{a)} we also overlay the mass size relation for 2$<$\,$z$\,$<$3 star-forming galaxies from \citet{Ward2024} and in panel \textit{b)} the relation between Asymmetry and Clumpiness identified by \citet{Conselice2003}. In panels \textit{c)} and \textit{d)} we show the  distributions of S\'ersic index ($n$) and  axis ratio ($b/a$) in the F444W band. For each distribution, we display the KS probability ($p_{ \rm KS}$) that \rev{each field sample and the SMG distributions} are drawn from the same parent population, where $p_{ \rm KS}$\,$<$\,0.001 suggests a significant difference in the distributions. We plot the bootstrap median and uncertainty for the two distributions at the top of each panel. The SMGs, on average, have comparable sizes and lower S\'ersic indexes  than \rev{both} field samples, whilst exhibiting similar axis ratios and clumpiness  with higher asymmetry.}
    \label{Fig:Morph_hist}
\end{figure*}

To summarise, while signs of potential dynamical disturbance are frequently seen in our SMG sample, we conclude that the majority of SMGs do not appear to be the result of late-stage, major mergers. This is in contrast with the situation for similar far-infrared luminous populations at $z$\,$\sim$\,0, where systems with far-infrared luminosities of $L_{\rm IR}$\,$\geq$\,10$^{12}$\,L$_\odot$ are frequently associated with late-stage major mergers \citep[e.g.,][]{Sanders1988,Farrah2001}, although there are claims that this fraction  declines at $z$\,$\sim$\,1 \citep[e.g.,][]{Kartaltepe2010}, with an increasing fraction of starburst galaxies appearing to be isolated systems \citep[e.g.,][]{Faisst2024}. It is also in contrast to early studies of SMGs with resolved CO kinematics \citep[e.g.,][]{Engel2010} as well near-infrared spectroscopic observations \citep[e.g.,][]{Zadeh2012,Drew2020} which suggested that the majority of SMGs are mergers, with distinct separated dust-detected components.  

\begin{figure*}
    \centering
    \includegraphics[width=\linewidth]{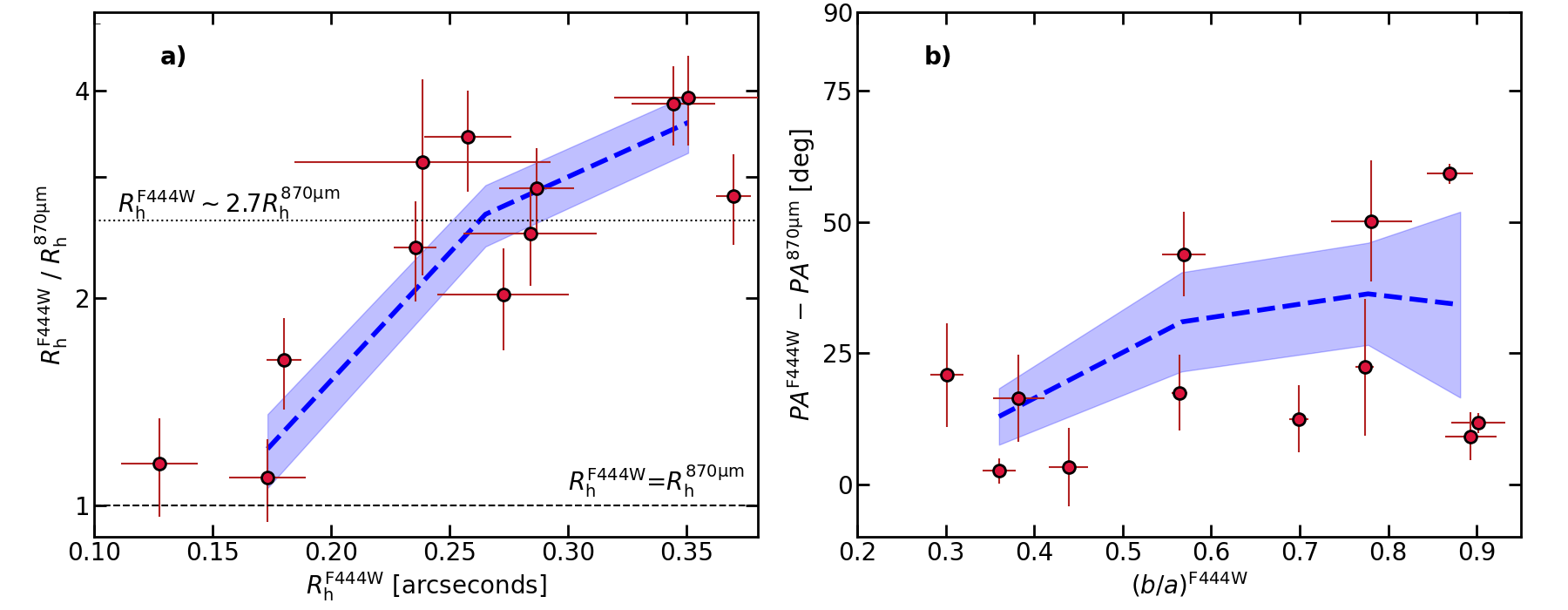}
    \caption{\rev{A comparison of the ALMA 870\um\ morphological properties to the NIRCam F444W morphologies for the subset of SMGs with high-resolution ALMA observations from \citet{Gullberg2019}. We compare (a) the half-light radii and (b) the position angle as a function of axis ratio, with a running median (blue line and shaded region) shown in each panel. On average we identify significantly smaller 870$\mu$m sizes ($R_{\rm h}^{\rm F444W}\,/\,R_{\rm h}^{\rm 870\um}\,$\,=\,2.7\,$\pm$\,0.4) but consistent axis ratios ($b/a_{\rm F444W}\,/\,b/a_{\rm 870\um}\,$\,=\,0.95\,$\pm$\,0.11) and position angles ($\rm PA_{\rm F444W}$\,$-$\,$\rm PA_{\rm 870\um}\,$\,=\,17\,$\pm$\,5$^{\circ}$), with a larger offset in position towards more circular axis ratios where the position angle is more degenerate.}}
    \label{fig:ALMA_compare}
\end{figure*}

\subsubsection{Sizes, s\'ersic indices and axis Ratios}

%half light size
We measure the half-light radius of the SMGs and field  galaxies in the F444W band, determining median values of $R_{\rm h}^{\rm SMG}$\,=\,2.70\,$\pm$\,0.23\,kpc and $R_{\rm h}^{\rm field}$\,=\,2.50\,$\pm$\,0.10\,kpc with 16\qth\,--\,84\qth\ percentile ranges of $R_{\rm h}^{\rm SMG}$\,=\,1.82\,--\,4.36\,kpc and $R_{\rm h}^{\rm field}$\,=\,1.64\,--\,3.76\,kpc respectively. \rev{For the massive field sample, we determine a median of $R_{\rm h}^{\rm M.\,field}$\,=\,2.42\,$\pm$\,0.15\,kpc with a 16\qth\,--\,84\qth\ percentile range of $R_{\rm h}^{\rm M.\,field}$\,=\,1.70\,--\,3.47\,kpc. This is comparable to the rest-frame 1.5\um\ size measured by \citet{Martorano2024} for massive $\rm \log_{10}(M_{\ast}[M_{\odot}])$\,=\,11.0\,--\,11.5 star-forming galaxies at 2.0\,$\leq$\,$z$\,$\leq$2.5 with a median size of $R_{\rm h}^{\rm 1.5\um}$\,=\,2.24\,kpc with a 16\qth\,--\,84\qth\ percentile range of $R_{\rm h}^{\rm 1.5\um}$\,=\,1.55\,--\,4.27\,kpc.}

To establish whether the SMGs and field  galaxies are drawn from the same underlying distribution, we perform a Kolmogorov-Smirnov (KS) test on the mass normalised distributions of half-light radius. To mass normalise the half-light radius distributions, we adopt the rest-frame optical mass size relation from \citet{Ward2024} at 2$<$\,$z$\,$<$3, noting that this is consistent with the trends seen in our samples. From the median stellar mass of the less activate field galaxies to the median stellar mass of the SMG sample, we derive an increase in half-light radius of 27\,$\pm$\,6\%, which we apply to the less massive field galaxies' half-light radii to account for the offset in stellar mass between the two samples. \rev{For the massive field sample we make a direct comparison to the SMGs}. For the non-Gaussian distributions of SMG and field galaxy morphological properties, we require $p_{ \rm KS}$\,$\leq$\,0.003 ($>$3\,$\sigma$) to confirm the populations are inherently different. We conclude the SMGs and both field galaxy samples have indistinguishable distributions of mass normalised half-light radius with $p_{\rm KS}$\,=\,0.06 \rev{for the less massive field sample and $p_{\rm KS}$\,=\,0.36 for the massive field sample}.

In Figure \ref{Fig:Morph_hist}a we show the correlation between stellar mass and F444W half-light radius for both field \rev{samples} and the SMGs. We overlay the rest-frame optical mass size relation for star-forming galaxies at 2$<$\,$z$\,$<$3 from \citet{Ward2024}. On average, the field  galaxies follow the trend identified by \citet{Ward2024}, whilst the SMGs, as indicated by the running median in Figure \ref{Fig:Morph_hist}a, exhibit marginally smaller F444W sizes for stellar masses above $\rm \log_{10}(M_\ast[M_{\odot}])$\,$>$\,10.5. We note however that the majority of galaxies used to define the mass size relation in \citet{Ward2024} have a stellar mass of $\log_{10}(M_\ast[M_{\odot}])$\,$<$\,10.5. In panel \textit{c)} of Figure \ref{Fig:Morph_hist}, we compare the distribution S\'ersic index for the field  sample to those of the SMGs, identifying more disc-like S\'ersic indices for the SMGs with a median of $n_{\rm F444W}$\,=\,1.10\,$\pm$\,0.10 and a 16\qth\,--\,84\qth\ percentile range of  $n_{\rm F444W}$\,=\,0.63\,--\,1.98 whilst the lower mass field galaxies have a higher median value of $n_{\rm F444W}$\,=\,1.85\,$\pm$\,0.07 with a 16\qth\,--\,84\qth\ percentile range of $n_{\rm F444W}$\,=\,0.88\,--\,4.35. \rev{The massive field sample has an even higher median S\'ersic index of $n_{\rm F444W}$\,=\,2.78\,$\pm$\,0.20 and a 16\qth\,--\,84\qth\ percentile range of  $n_{\rm F444W}$\,=\,1.51\,--\,4.89}. Applying a KS test, we identify \rev{that the SMGs are distinct compared to either the field or the massive field distributions with $p_{\rm KS}$\,$<$\,0.001 for both}\footnote{We note that the exclusion of the candidate mergers does not affect this result.}.

%axis ratio
From our non-parametric analysis, we derive a median axis ratio of $b/a$\,=\,0.55\,$\pm$\,0.03 for the SMGs and a 16\qth\,--\,84\qth\, percentile range of $b/a$\,=\,0.37\,--\,0.82 in the NIRCam F444W band. This is comparable to the axis ratio distribution for 870\um\ dust continuum in SMGs with $b/a$\,=\,0.63\,$\pm$\,0.02 \citep{Gullberg2019}. For the field  galaxies, we derive a median F444W axis ratio of $b/a$\,=\,0.53\,$\pm$\,0.01 and 16\qth\,--\,84\qth\, percentile range $b/a$\,=\,0.31\,--\,0.77, \rev{whilst for the massive field sample we derive a median $b/a$\,=\,0.64\,$\pm$\,0.03 and 16\qth\,--\,84\qth\, percentile range $b/a$\,=\,0.35\,--\,0.84}. The distribution of axis ratios, as shown in Figure \ref{Fig:Morph_hist}d, is comparable to that of the SMGs with $p_{\rm KS}$\,=\,0.23 \rev{for the field sample and $p_{ \rm KS}$\,=\,0.72 for the massive field sample}. 

\begin{figure*}
    \includegraphics[width=\linewidth]{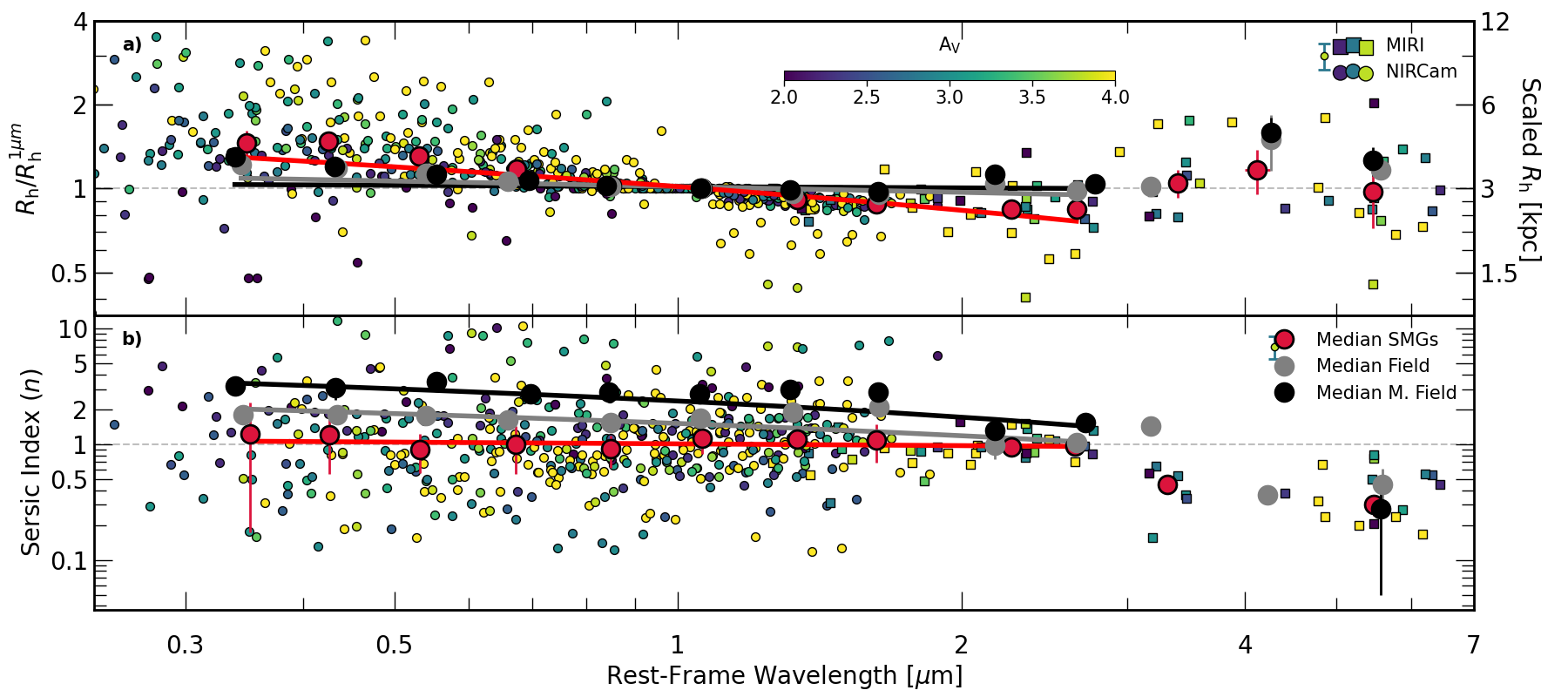}

     \caption{\textit{a)} The growth curve half-light radius ($R_{\rm h}$) of the SMGs, normalised by the rest-frame 1\um\ half-light radius of the respective galaxy as a function of rest-frame wavelength and colour-coded by the galaxy integrated \av. We show individual measurements for SMGs, and a representative error bar, as well as the running median for the field sample \rev{(grey points and line) and massive field sample (black points and line)}. \rev{On the right-hand axis we show the normalised half-light radius scaled by the median rest-frame 1\um\ size ($R_{\rm h}^{1\um}$\,=\,3.1\,kpc) for the SMGs.} \textit{b)} S\'ersic index derived from the \texttt{\sc{galfitm}} parametric fitting as a function of rest-frame wavelength. We indicate the NIRCam measurements  by circles and MIRI measurements by squares. We show a running median, and standard error ($\sigma/\sqrt{n}$) in fixed logarithmic bins of wavelength for the SMGs and field  galaxies. In both panels, we plot the parametric fit for the field (solid grey line) and SMGs (solid red line). We find that the SMGs, on average, have a stronger size variation with wavelength, becoming more compact at longer wavelengths quicker than either field sample, whilst exhibiting lower S\'ersic index at all wavelengths.}
    \label{Fig:Morph_wav}
\end{figure*}

\subsubsection{Residuals, clumpiness and asymmetry}

Whilst the S\'ersic profiles indicate more disc-like surface brightness distributions in F444W for the SMGs, it is the deviations from the parametric light profiles, as quantified by the RFF parameter, that encodes the detailed morphological properties of each galaxy. In Figure \ref{Fig:RFF}, we show examples of the NIRCam F444W S\'ersic model residuals, and the derived RFF values, for both SMGs and field  galaxies. Whilst some galaxies exhibit smooth residuals, and low RFF values, several galaxies (both SMGs and field) display complex morphologies with multiple clumps, spiral arms or bright compact point sources. For the SMGs we identify a median RFF value in the F444W band of RFF$^{\rm F444W}$\,=\,10.2\,$\pm$\,0.5 with a 16\qth\,--\,84\qth\ percentile range of RFF$^{\rm F444W}$\,=\,5.9\,--\,16.3 whilst the field  galaxies have a median value of RFF$^{\rm F444W}$\,=\,8.5\,$\pm$\,0.2 with a 16\qth\,--\,84\qth\ percentile range of RFF$^{\rm F444W}$\,=\,4.5\,--\,17.2. \rev{The massive field galaxies have a median RFF value in the F444W band of RFF$^{\rm F444W}$\,=\,7.0\,$\pm$\,0.63 with a 16\qth\,--\,84\qth\ percentile range of RFF$^{\rm F444W}$\,=\,4.4\,--\,11.85.} We compare the distributions of RFF for SMGs and field samples finding a KS-statistic of $p_{\rm KS}$\,=\,0.002 \rev{for the lower mass field sample, and $p_{\rm KS}$\,$<$0.001 for the high mass field. This indicates that the larger, less massive, field sample is not distinguishable at $>$3$\sigma$ level to the SMGs, whilst the massive field sample, with lower RFF values, is distinct from the SMGs.}

Although the 870\um\ ALMA observations of the majority of the SMGs in our sample are not high enough resolution (FWHM\,$\leq$\,0\farcs{2}) to identify these structural features, high-resolution sub-millimetre studies have identified spiral arms, bars and star-forming rings embedded in exponential dust disks \citep[e.g.][]{Hodge2019,Amvrosiadis2024}. For a subset of the SMGs (15/80), however, 0\farcs{2} resolution ALMA 870\um\ observations are available as discussed by \citet{Gullberg2019}. These resolved observations are insufficient to identify complex dust structures, but the resolved 870\um\ emission's morphology provides insight into any underlying structural features. For these 15 SMGs we first compare the ALMA 870\um\ size measured in the $uv$-plane to the F444W growth curve size\rev{, as shown in Figure \ref{fig:ALMA_compare}a}. We estimate a median ratio of $R_{\rm h}^{\rm F444W}\,/\,R_{\rm h}^{\rm 870\um}\,$\,=\,2.7\,$\pm$\,0.4 with a 16\qth\,--\,84\qth\ range of $R_{\rm h}^{\rm F444W}\,/\,R_{\rm h}^{\rm 870\um}\,$\,=\,1.5\,--\,3.5. As identified by previous studies the far-infrared emission is much more compact than the stellar emission at shorter wavelengths \citep[e.g.][]{Simpson2015,Lang2019,Gullberg2019, Chen2022}. We perform a Spearman rank test on the correlation between F444W/870\um\ size ratio and the \av\ of the galaxy, identifying no significant correlation ($p_{\rm SR}$\,=\,0.33). However, a larger sample is required to robustly confirm the lack of correlation.

To investigate the presence of complex dust structures we compare the 870\um\ axis ratio and position angle to that derived for the F444W emission. For the 870\um\ emission we adopt the free S\'ersic fitting parameters from \citet{Gullberg2019}. We identify median ratios of $b/a_{\rm F444W}\,/\,b/a_{\rm 870\um}\,$\,=\,0.95\,$\pm$\,0.11 and, \rev{as shown in Figure \ref{fig:ALMA_compare}b}, a median offset in axis ratio of $\rm PA_{\rm F444W}$\,$-$\,$\rm PA_{\rm 870\um}\,$\,=\,17\,$\pm$\,5\,$^{\circ}$, indicating good agreement in the alignment of the near-infrared (rest-frame $\simeq$\,1\um) and far-infrared (rest-frame $\simeq$\,250\um) emission, suggesting they are tracing the same underlying structure, in contrast to the UV/optical and far-infrared offsets identified in previous multi-wavelength high-resolution studies \citep[e.g.,][]{Chen2015,Rivera2018,Hodge2019}.

%statmorph
In Figure \ref{Fig:Morph_hist}b, we show the correlation between asymmetry ($A$) and clumpiness ($S$) in the NIRCam F444W band as derived from \texttt{\sc{statmorph}} and over plot the relation between $A$ and $S$ derived by \citet{Conselice2003}\footnote{The relation is corrected for the different definitions of clumpiness given by \citet{Lotz2004} and \citet{Conselice2003}}. We find that the SMGs and field  galaxies have similar clumpiness with a median value of $S_{\rm F444W}^{\rm SMGs}$\,=\,0.02$^{+0.03}_{-0.02}$ and $S_{\rm F444W}^{\rm field }$\,=\,0.03$^{+0.04}_{-0.04}$ respectively with a $p_{ \rm KS}$\,=\,0.02. \rev{The massive field sample has $S_{\rm F444W}^{\rm M. Field}$\,=\,0.03$^{+0.05}_{-0.01}$, with a $p_{ \rm KS}$\,=\,0.01. This indicates that both field and massive field samples are broadly consistent with being drawn from similar distributions to the SMGs.} The asymmetry in the F444W band for the SMGs is higher with a median value of $A_{\rm F444W}^{\rm SMGs}$\,=\,0.13$^{+0.09}_{-0.06}$ compared to the field  galaxies ($A_{\rm F444W}^{\rm field }$\,=\,0.08$^{+0.12}_{-0.06}$) and a $p_{ \rm KS}<$\,0.001. \rev{The massive field sample has median value of $A_{\rm F444W}^{\rm M. field}$\,=\,0.07$^{+0.06}_{-0.03}$ with a $p_{ \rm KS}<$\,0.001 indicating the field galaxies and massive field galaxies asymmetry distributions are distinct from the SMGs.} 

The correlation between asymmetry and clumpiness highlights two distinct regions. At constant asymmetry, there are two populations, one at low clumpiness ($S$\,$<$\,0.04)  and one at high clumpiness ($S$\,$>$\,0.04), where the majority of the SMGs and field  galaxies have low clumpiness with smooth, symmetric light distributions whilst those galaxies at high clumpiness show signs of concentrated star-forming regions. The SMGs at $S$\,$>$\,0.04 are more compact in the F444W filter with a median size of $R_{\rm h}$\,=\,1.72\,$\pm$\,0.14\,kpc compared to the SMGs with lower clumpiness ($S$\,$<$\,0.04) with a median size of $R_{\rm h}$\,=\,2.96\,$\pm$\,0.25\,kpc, whilst exhibiting comparable morphological and {\sc{magphys}} derived properties. We note that several studies have highlighted the uncertainty introduced in non-parametric morphological measurements when the Petrosian radius of the galaxy ($R_{\rm p}$) is comparable to the HWHM of the observations \citep[e.g.,][]{Yu2023,Ren2024}, especially when the clumpiness parameter is measured between 0.25$R_{\rm p}$ and 1.5$R_{\rm p}$, see \citet{Lotz2004} for details. This suggests that the non-parametric measurements for these compact galaxies maybe more uncertain than for the more extended sources.

\begin{figure*}
    \centering
    \includegraphics[width=\linewidth]{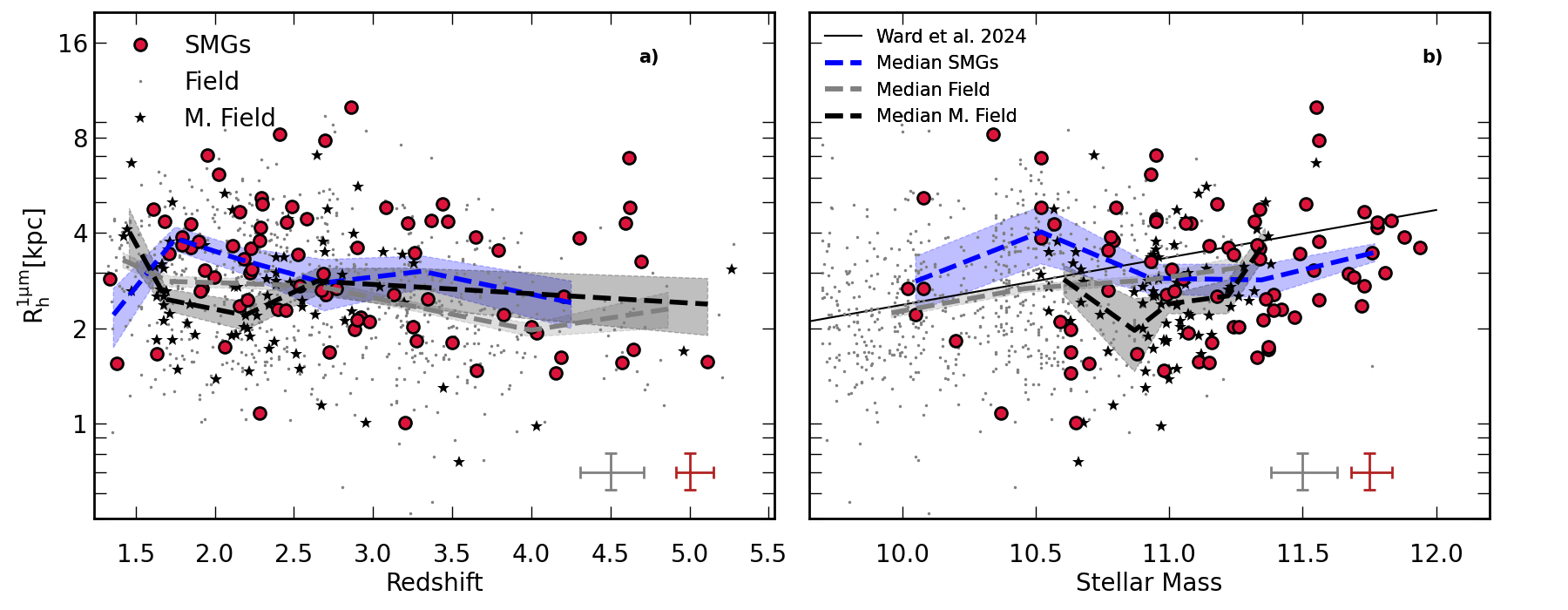}
    \caption{The rest-frame 1\um\ half-light radius for SMGs ({\it{red circles}}), less massive field galaxies ({\it{grey dots}}) \rev{and massive field galaxies ({\it{black stars}})} as a function of  ($a$) redshift and ($b$) stellar mass . In each panel, we show a running median for SMGs (dashed blue line), field galaxies (grey dashed line) \rev{and massive field galaxies (black dashed line}) as well as a representative uncertainty. We identify a consistent reduction in rest-frame 1\um\ size with increasing redshift in both field samples and the SMG sample, as well as a consistent mass size relation compared to that seen in Figure \ref{Fig:Morph_hist}. This indicates that the morphology variation with wavelength identified in both field samples and the SMGs in Figure \ref{Fig:Morph_wav} is not driven by K-correction effects.}
    \label{fig:1m_sizes}
\end{figure*}

\subsection{Wavelength dependent morphology}

\begin{figure*}
    \includegraphics[width=\linewidth]{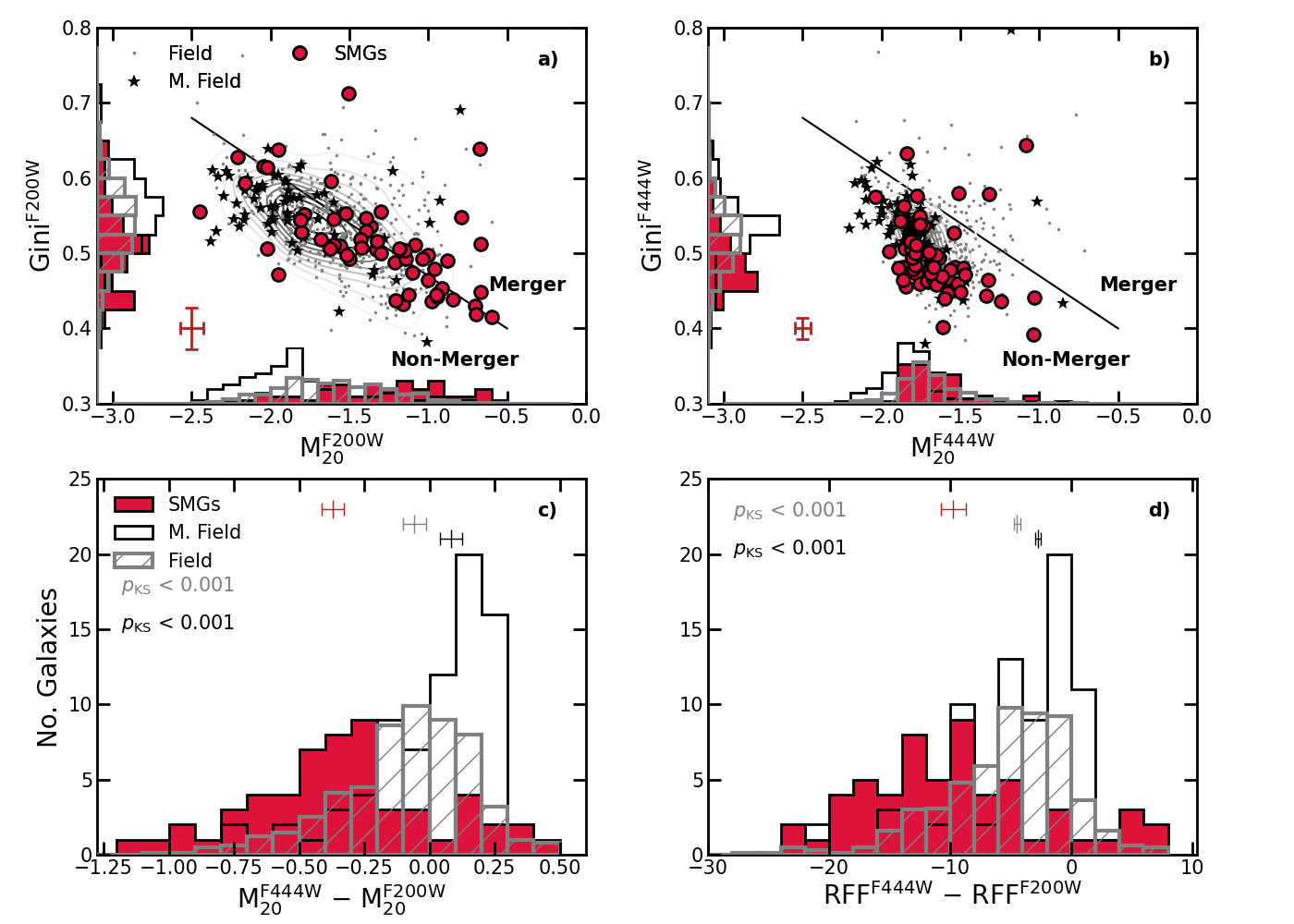}

     \caption{We show the Gini\,--\,M$_{20}$ relation for the \rev{less active} field galaxies (grey dots and contours), \rev{massive field galaxies (black stars)} and SMG sample (red circles), in the \textit{a)} F200W  and \textit{b)} F444W bands as well as histograms showing the distribution of each parameter on each axis, \rev{with the low mass field scaled down by a factor 10}. The solid-black line indicates the boundary line between mergers and non-mergers as defined in \citet{Lotz2008}. Both SMGs and field  galaxies scatter about the line in the F200W band whilst the majority lie in the non-merger region of the equivalent F444W plot. In panels  \textit{c)} and  \textit{d)} we show the distributions of M$_{\rm 20}^{\rm F444W}$\,$-$ \,M$_{\rm 20}^{\rm F200W}$ and  RFF$^{\rm F444W}$\,$-$ \,RFF$^{\rm F200W}$. For each distribution, we indicate the KS-statistic showing that the SMGs have a stronger variation in morphology with wavelength with higher RFF and  M$_{\rm 20}$ values at 2\um\ than 4\um\ indicating more structure in the residuals and disturbed morphologies.}
     \label{Fig:Gini}
\end{figure*}

The preceding analysis of the  SMGs and field  galaxies stellar (near-infrared) morphologies has revealed broad similarities between the two populations. Many previous studies suggest the extreme properties of SMGs originate from merger-driven events \citep[e.g.,][]{Smail1998,Swinbank2010,Aguirre2013}. We do not expect the morphological signatures of these mergers to show wavelength dependence in contrast to the effects of dust. Thus, by quantifying the morphological variation with wavelength in the field and SMG sample, we can test the applicability of the merger-driven scenario. 

For the SMGs and field  galaxies we measure a median F200W half-light radii of $R_{\rm h,F200W}^{\rm SMG}$\,=\,4.0\,$\pm$\,0.3\,kpc and $R_{\rm h,F200W}^{\rm field}$\,=\,2.8\,$\pm$\,0.1\,kpc respectively. Compared to F444W, for the SMGs this represents a 48\,$\pm$\,0.17\% reduction in size between 2\um\ and 4\um\, whilst for the field  galaxies the reduction in size is only 12\,$\pm$\,6\%, although we note this comparison is not done at fixed stellar mass it highlights the variation of morphology with wavelength for the two samples. \rev{The massive field sample has a median F200W half-light radii of $R_{\rm h,F200W}^{\rm M. field}$\,=\,2.9\,$\pm$\,0.2\,kpc which corresponds to a 20\,$\pm$\,11\% reduction in size.}  To further investigate the variation in the SMGs and field  galaxy morphology as a function of wavelength, in Figure \ref{Fig:Morph_wav} we plot the half-light radius, normalised by the rest-frame 1\um\ half-light radius, and S\'ersic index against the rest-frame wavelength probed by the NIRCam and MIRI observations. 

Below 1\um, for SMGs in Figure \ref{Fig:Morph_wav} there is increased spread in half-light radius with an apparent correlation with dust content as quantified by \av. At rest-frame wavelengths greater than 3\um\, where we use the MIRI F770W and F1800W bands to quantify the morphology, we identify a decrease in the median S\'ersic index of both field and SMGs to $n$\,$\sim$\,0.4 and an increase in the half-light radius. We suspect that this trend is driven by the larger beam size of the MIRI observations with FWHM\,=\,0\farcs{27} and \,0\farcs{59} respectively for the F770W and F1800W filters, in combination with the shallower depth compared to the NIRCam observations, resulting in a smoother, less-structured, light distribution. In addition, in the rest-frame near-infrared ($>$\,3\um) any AGN component that is present in the galaxy starts to be more prominent over the stellar emission, with point-source ($n$\,$\sim$\,0.5) morphology.

To quantify the field  galaxy and SMGs size evolution with wavelength, we perform fits to the median values shown in Figure \ref{Fig:Morph_wav} up to rest-frame 3\um\ using an orthogonal distance relation (ODR) algorithm that takes into account the uncertainties on the median values in both wavelength and size. We define the function as,
\begin{equation}
     \frac{R(\lambda)}{R_{1\um}}\,=\,\frac{d(R_{\rm h}/R_{1\um})}{d\lambda}\log_{10}(\lambda),%R_0+
\end{equation}
where $\lambda$ is the rest-frame wavelength probed by the observations. For the SMGs we derive a slope of $\frac{d(R_{\rm h}/R_{1\um})}{d\lambda}$\,=\,$-$0.60\,$\pm$\,0.09 whilst for the field galaxies we estimate $\frac{d(R_{\rm h}/R_{1\um})}{d\lambda}$\,=\,$-$0.15\,$\pm$\,0.07 \rev{and the massive field galaxies $\frac{d(R_{\rm h}/R_{1\um})}{d\lambda}$\,=\,$-$0.04\,$\pm$\,0.09.}
This indicates the SMGs have stronger variation in size with wavelength, becoming more compact faster than \rev{either field sample}, as highlighted by the best-fit solution plotted in Figure \ref{Fig:Morph_wav}. However, this trend may be driven by centrally concentrated dust inflating the rest-frame optical sizes of the SMGs resulting in a stronger observed variation with wavelength. To ensure size ($R_{\rm h}$) variation with wavelength identified in the SMGs and field galaxies in Figure \ref{Fig:Morph_wav} is not driven by the redshift evolution of the galaxies, in Figure \ref{fig:1m_sizes}, we correlate the rest-frame 1\um\ size of the galaxies with their redshift. On average we identify a consistent redshift evolution in the near-infrared sizes of the SMGs and \rev{both field samples}, indicating that redshift dependent K-corrections are not driving the morphological wavelength variation.

\begin{figure*}
    \includegraphics[width=\linewidth]{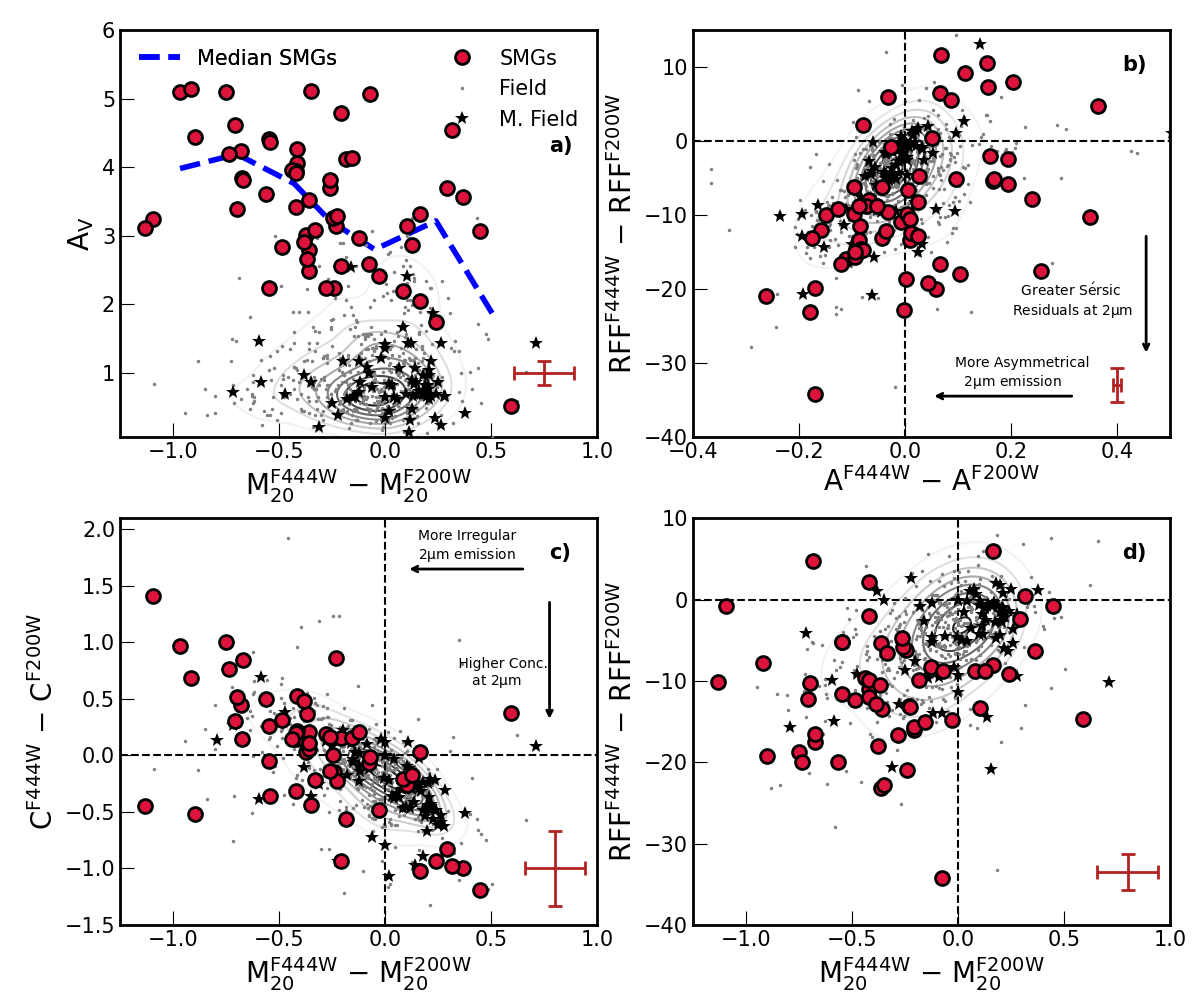}
     \caption{A comparison of the 2\um\ and 4\um\ morphological properties of the SMGs and \rev{both field samples}. We compare the difference in  M$_{\rm 20}$ parameter between the  F444W and F200W bands  (M$^{\rm F444W}_{\rm 20}$\,$-$ \,M$^{\rm F200W}_{\rm 20}$) with dust attenuation (\av) in panel \textit{a)}. In panel \textit{b)}, we further compare the difference in RFF between the F444W and F200W bands ($\rm RFF^{F444W}-RFF^{F200W}$) with the difference in Asymmetry ($A^{\rm F444W}-A^{\rm F200W}$). Finally  we plot the difference in M$_{\rm 20}$ between F444W and F200W (M$_{\rm 20}^{\rm F444W}$\,$-$\,M$_{\rm 20}^{\rm F200W}$) with the difference in Concentration (C$^{\rm F444W }_{\rm}$\,$-$\,C$^{\rm F200W}$) (panel \textit{c)} and  the difference in RFF  (panel \textit{d)}. All four panels indicate the SMGs are much more structured than \rev{either field population}, with less uniform light distributions in the F200W band compared to F444W, in contrast,  the field galaxies show minimal wavelength variation.}
    \label{Fig:Non-para_wav}
\end{figure*}

Furthermore Figure \ref{Fig:Morph_wav} shows that the SMGs, on average, have a lower S\'ersic index than the field galaxies at all wavelengths. To quantify the evolution of S\'ersic index with wavelength we fit a similar relation to that used to quantify the size evolution of the form,
\begin{equation}
  n(\lambda)\,=\,n_0+\frac{dn}{d\lambda}\log_{10}(\lambda),
\end{equation}
up to a rest-frame wavelength of 3\um. The median S\'ersic index of the SMGs have an $\frac{dn}{d\lambda}$\,=\,$-$0.12\,$\pm$\,0.11 and $n_0$\,=\,1.00\,$\pm$\,0.04, whilst the field galaxies exhibit strong variation with wavelength, but at higher average S\'ersic index ($\frac{dn}{d\lambda}$\,=\,$-$1.10\,$\pm$\,0.23 and $n_0$\,=\,1.51\,$\pm$\,0.10). \rev{For the massive field galaxies we measure $\frac{dn}{d\lambda}$\,=\,$-$2.16\,$\pm$\,0.81 and $n_0$\,=\,2.37\,$\pm$\,0.32.} Thus indicating on average, the SMGs have light distributions closer to pure disk-like exponentials. 

To further investigate the variation of morphology with wavelength in the SMG and field  samples, in Figures \ref{Fig:Gini}a \& b, we show the Gini\,--\,M$_{20}$ relation which has been used as a crude late-stage merger indicator \citep{Lotz2008, Liang2024,Polletta2024} in the F200W and F444W bands. We select the F444W band as this most closely traces the stellar morphology of the galaxies, whilst F200W band corresponds to the rest-frame $V$-band for the median redshift of the samples, at a similar wavelength to the $R$-band where the Gini\,--\,M$_{20}$ ``merger'' and ``non-merger'' boundary was calibrated by \citet{Lotz2008}. At 2\um\, (rest-frame $V$-band) both field  and SMG samples scatter about the boundary between ``merger'' and ``non-merger'', with 40\,$\pm$\,5\% of the SMGs lying in the merger region. Although the SMGs are offset relative to the field populations with higher M$_{20}$ and lower Gini values. At 4\um\, (rest-frame $z$-band) almost all SMGs (92\,$\pm$\,2\%) and field galaxies lie in the ``non-merger'' region of the relation, with consistent Gini and M$_{20}$ values. Between 2\um\ and 4\um\, the relative shift in the Gini\,--\,M$_{20}$ parameters is larger for the SMGs than for the field samples.

To understand the origin of this shift in the Gini-M$_{20}$ plane, in Figure \ref{Fig:Gini}c we show the difference between the M$_{\rm 20}$ parameter in the F444W and F200W bands. This plot highlights that the SMGs have higher M$_{20}$ in the F200W than F444W band which indicates SMGs have less concentrated F200W morphologies whilst both the field \rev{and massive field} galaxies have more similar M$_{\rm 20}$ values between the two bands. The significance of the difference between the field \rev{samples} and SMGs is quantified by the KS-statistic value of $p_{\rm ks}$\,$<$\,0.001. To highlight the reduction in structured emission between the F200W and F444W bands, in Figure \ref{Fig:Gini}d we show the difference between RFF in the two bands for the SMGs and field  galaxies. On average the SMGs exhibit a greater difference in RFF across bands, with more significant F200W residuals, with a median value of RFF$^{\rm F444W}$\,$-$\,RFF$^{\rm F200W}$\,=\,$-$10.0\,$\pm$\,1.0 with a 16\qth\,--\,84\qth\ percentile range of RFF$^{\rm F444W}$\,$-$\,RFF$^{\rm F200W}$\,=\,$-$2.0 to $-$17.4 whilst for the field  galaxies the median is RFF$^{\rm F444W}$\,$-$\,RFF$^{\rm F200W}$\,=\,$-$4.6\,$\pm$\,0.3 with a 16\qth\,--\,84\qth\ percentile range of RFF$^{\rm F444W}$\,$-$\,RFF$^{\rm F200W}$\,=\,$-$0.2 to $-$11.6. \rev{The massive field galaxies have RFF$^{\rm F444W}$\,$-$\,RFF$^{\rm F200W}$\,=\,$-$3.2\,$\pm$\,0.8 with a 16\qth\,--\,84\qth\ percentile range of RFF$^{\rm F444W}$\,$-$\,RFF$^{\rm F200W}$\,=\,$-$0.15 to $-$9.5.} This is additional evidence of the dust obscuration at shorter wavelengths in the SMGs, resulting in deviations from a simple light profile. This comparison suggests the SMGs have more complex morphologies at bluer wavelengths, which is likely to be linked to their intense star formation rates and the presence of highly structured dust in the SMGs, attenuating the shorter wavelength light.

To identify the physical mechanism driving the increase in structured emission identified in the SMGs (as quantified by the RFF and M$_{20}$) as a function of wavelength, in Figure \ref{Fig:Non-para_wav} we compare the difference in morphological parameters measured between the F444W and F200W NIRCam bands for the SMGs and field populations. In Figure \ref{Fig:Non-para_wav}a we compare the dust content, as quantified by the \av\ from SED fitting (Section \ref{subsec:phot}), with M$_{20}^{\rm F444W}$\,$-$\,M$_{20}^{\rm F200W}$. As shown in Figure \ref{Fig:SED_hist}, \rev{both field samples} have significantly lower \av\ and less morphological variation with wavelength, with a median M$_{20}^{\rm F444W}$\,$-$\,M$_{20}^{\rm F200W}$\,=\,$-$0.06\,$\pm$\,0.01 \rev{for the field sample and M$_{20}^{\rm F444W}$\,$-$\,M$_{20}^{\rm F200W}$\,=\,0.09\,$\pm$\,0.03 for the massive field sample.} The SMGs on the other hand, with higher \av\, indicate a greater disparity between the F200W and F444W light distribution with a median value of M$_{20}^{\rm F444W}$\,$-$\,M$_{20}^{\rm F200W}$\,=\,$-$0.34\,$\pm$\,0.04. To quantify the trend between \av\, and  M$_{20}^{\rm F444W}$\,$-$\,M$_{20}^{\rm F200W}$ we perform an ODR fit to the running median shown in Figure \ref{Fig:Non-para_wav} of the form,
\begin{equation}
 {\rm A}_{V}\,=\,{\rm A}_{V,0}+\alpha ({\rm M}_{20}^{\rm F444W}-{\rm M}_{20}^{\rm F200W}),
\end{equation}
We determine for the SMGs best-fit parameters of $\alpha$=\,$-$1.0\,$\pm$0.3 and \av$_{,0}$\,=\,3.1\,$\pm$\,0.1, indicating a strong negative correlation for the SMGs, whereby more dust obscured (higher \av) SMGs, have a more negative M$_{20}^{\rm F444W}$\,$-$\,M$_{20}^{\rm F200W}$, indicating less concentrated light profiles in the F200W band compared to the F444W.

To analyse the connection to other morphological parameters, in Figures \ref{Fig:Non-para_wav}b \& d we plot RFF$^{\rm F444W}$\,$-$\,RFF$^{\rm F200W}$ against the difference in Asymmetry (A$^{\rm F444W}$\,$-$\,A$^{\rm F200W}$) and  M$_{20}^{\rm F444W}$\,$-$\,M$_{20}^{\rm F200W}$, whilst in Figure \ref{Fig:Non-para_wav}c we compare the difference in Concentration (C$^{\rm F444W}$\,$-$\,C$^{\rm F200W}$) with M$_{20}^{\rm F444W}$\,$-$\,M$_{20}^{\rm F200W}$. All three relations indicate that on average the SMGs in the F200W band are more asymmetric and less concentrated, with more in-homogeneous light distributions leading to larger residuals to a single S\'ersic fit  at 2\um\ as compared to their 4\um\ morphologies. 

\section{Discussion}\label{Sec:Disc} 

\subsection{Does structured dust drive the differences between SMGs and field galaxies?}

The morphological properties of the SMGs in the NIRCam F444W band are comparable to the field  galaxies, with similar sizes at a fixed mass, slightly lower S\'ersic indices and higher asymmetry (Figure \ref{Fig:Morph_hist}). The F444W filter traces the rest-frame near-infrared ($\approx$1\um) out to $z$\,$\sim$\,3.5, and thus is less affected by dust obscuration and recent star formation than shorter bands in most of our sample. A greater difference between SMGs and field samples is identified at bluer wavelengths.

\begin{figure*}
\centering
\begin{subfigure}{.49\linewidth}
  \centering
 \includegraphics[width=\linewidth]{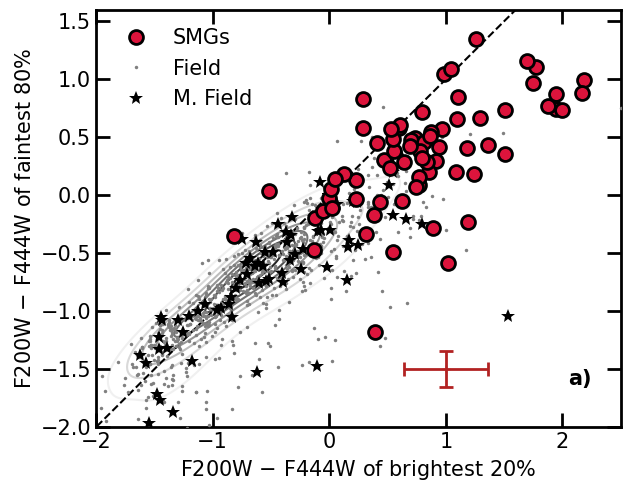}
\end{subfigure}%
\begin{subfigure}{.51\linewidth}
  \centering
  \includegraphics[width=\linewidth]{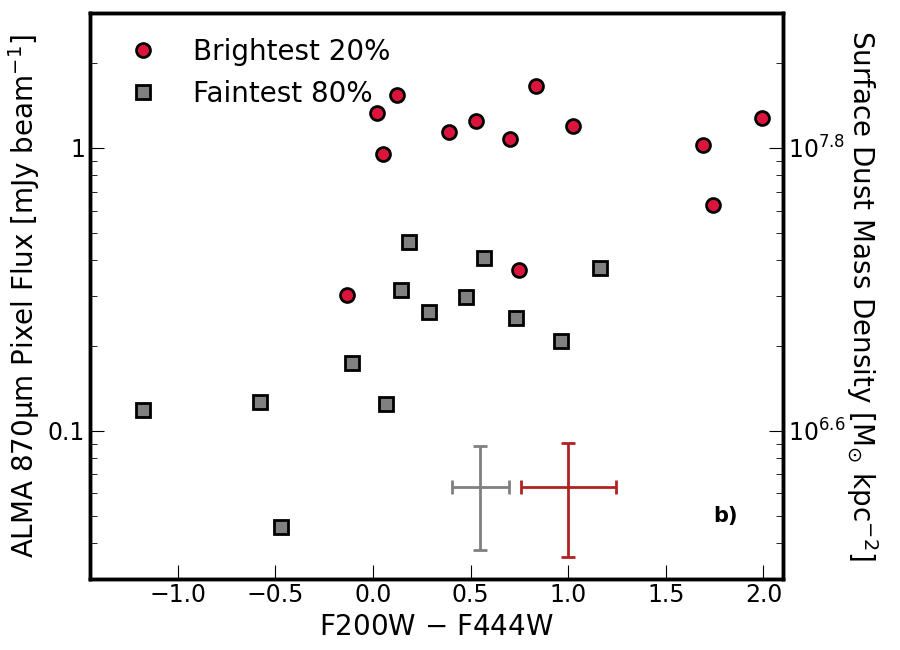}
\end{subfigure}
 \caption{\textit{a)} We plot the average F200W\,$-$\,F444W colour of the faintest pixels (below 80\qth\ percentile of flux) inside 0\farcs{5} elliptical apertures in the F444W image for the SMGs and \rev{field samples}, as a function of the average colour for the top 20 per cent brightest F444W pixels. \rev{Both field samples} exhibit almost uniform colours between faint and bright regions, with a small offset to the equality line of 0.12\,$\pm$\,0.03 mag (in the sense that the brighter regions are redder). The SMGs are both generally redder in fainter regions and significantly redder (1.11\,$\pm$\,0.09 mag) in their brightest regions. In panel \textit{b)} we show the average F200W\,$-$\,F444W colour of the ``Brightest 20\%'' and ``Faintest 80\%'' pixels as a function of ALMA pixel brightness. On the right-hand axis we show the surface dust mass density derived using the 870\um\ flux to dust mass relation from \citet{Dud2020} at the median redshift of the sample ($z\,=\,2.7$).  The brightest regions in the near-infrared (NIRCam F444W), which have the reddest F200W\,$-$\,F444W colour, correspond to the brightest regions in the rest-frame far-infrared (ALMA 870\um).}
 \label{Fig:M20_colours}
\end{figure*}

By investigating the variation M$_{20}$ parameter and RFF between the NIRCam F444W band ($\lambda_{\rm rest}\approx$\,1\um) and F200W band ($\lambda_{\rm rest}\approx$\,0.5\um), Figures \ref{Fig:Gini} and \ref{Fig:Non-para_wav} highlight that the SMGs have more structured light profiles at shorter wavelengths, that results in larger residuals to single S\'ersic fits. This morphological variation with wavelength was also highlighted by \citet{Chen2016} for SMGs and \citet{Nedkova2024} for SFGs, the latter identifying that the most massive (M$_{\ast}$\,$\gtrsim$10$^{10}$M$_{\odot}$) disc galaxies in their sample at 0.5\,$<$\,$z$\,$<$\,3 have larger half-light radii in the rest-frame ultra-violet than optical, which they attribute to the significant dust attenuation in the central regions of the galaxies compared to the outskirts. A similar connection between with inferred galaxy morphology and dust content was identified in the Illustris -- The Next Generation \citep[TNG50;][]{Pillepich2019} simulation by \citet{Popping2022}. They suggested that the observed $H$-band to 870\um\ size ratio increases towards higher redshift (hence bluer rest-frame sampling) due to dust attenuating the central regions of galaxies, resulting in larger half-light radii in the $H$-band. The impact of dust on size measurements has also been quantified in the First Light And Reionisation Epoch Simulations \citep[FLARES;][]{Lovell2021a,Vijayan2021}, where intrinsically massive compact galaxies appear significantly more extended when the effects of dust are taken into account \citep{Roper2022,Roper2023}. 

To isolate the contribution of dust to the SMGs morphological evolution as a function of wavelength, we examine the F200W\,$-$\,F444W pixel colours. In particular, given the M$_{20}^{\rm F444W}$\,$-$\,M$_{20}^{\rm F200W}$ parameter indicates the greatest morphological difference between field galaxies and SMGs, we calculate the average colour of the brightest 20 percent of pixels in the F444W image inside the 0\farcs{5} elliptical apertures of each galaxy (Section \ref{subsec:phot}). In Figure \ref{Fig:M20_colours}a, for the SMGs and \rev{ both field samples}, we compare this ``Brightest 20\%'' colour with the average colour of the fainter regions inside the same aperture (``Faintest 80\%'' F200W\,$-$\,F444W). Galaxies exhibiting no strong colour gradients between bright and faint regions will have consistent ``Brightest 20\%'' and ``Faintest 80\%'' colours, lying close to the one-to-one line. The majority of the field galaxies, \rev{as well as the massive field galaxies}, as shown in Figure \ref{Fig:M20_colours}a, fall into this regime, with on average the brightest regions of the field galaxies being 0.12\,$\pm$\,0.03 mag, redder than the faintest. 

For the SMGs the F200W\,$-$\,F444W pixel colour of the fainter regions that comprise the bulk of the galaxies (``Faintest 80\%'') indicate on average redder colours, with a median colour of 0.40\,$\pm$\,0.06\,mag compared to $-$0.74\,$\pm$\,0.02\,mag for the field  galaxies \rev{and $-$0.67\,$\pm$\,0.08\,mag for the massive field galaxies}. For the ``Brightest 20\%'' pixel colour, again the SMGs indicate significantly redder colours with a median value of 0.77\,$\pm$\,0.06\,mag compared to $-$0.66\,$\pm$\,0.02\,mag for the field  galaxies \rev{and $-$0.66\,$\pm$\,0.03\,mag for the massive field galaxies}. 

To understand whether the reddest (and brightest) regions of the SMGs are physically associated with the massive dust reservoirs, and thus 870\um\ emission, of the galaxies we analyse high-resolution ALMA 870\um\ observations for a sub-sample of 15 SMGs with ALMA maps at $\leq$0\farcs{2} FWHM resolution from \citet{Gullberg2019}. We resample the ALMA maps to match the 0\farcs{04} pixel scale of the \jwst\ NIRCam F444W imaging. We then extract the same average sub-millimetre flux density of the regions corresponding to the ``Brightest 20\%'' and ``Faintest 80\%'' F444W pixel as before for each SMG. In Figure \ref{Fig:M20_colours}b, we correlate this pixel colour with the ALMA 870\um\ pixel brightness for the ``Brightest 20\%'' and ``Faintest 80\%'' pixels. We identify a strong correlation between F200W\,$-$\,F444W pixel colour and 870\um\ surface brightness, demonstrating that the redder colours that originate from the brightest regions of the SMGs are spatially correlated with the brightest regions of the 870\um\ emission and thus high dust column density regions. 

The high dust column density of the SMGs, as shown in Figure \ref{Fig:M20_colours}, implies that the optical to near-infrared emission of the SMGs is strongly attenuated. Consequently, physical properties estimated from the SED fitting (e.g. \av) may be uncertain due to more heavily obscured regions of the galaxies being undetected, a situation that will be even worse when relying solely on UV to near-infrared photometry and no energy-balance assumptions. We can derive the expected $V$-band extinction (\av) given the median dust mass ($\rm \log_{10}(M_d[M_{\odot}])$\,=\,8.9\,$\pm$\,0.2) derived from SED fitting (Sec \ref{subsec:phot}) and compare this to the \av\ derived by \texttt{\sc{magphys}} from fitting the optical to near-infrared SED. 

As shown by \citet{Guver2009}, the column density of hydrogen (in cm$^{-2}$) can be approximated from the \av\ (in mag) as follows,
\begin{equation}
     N_{H}\,=\,2.21\,\times 10^{21}\,{\rm A}_{V}\,
\end{equation}
Adopting a dust-to-gas ratio of $\delta$\,=\,63\,$\pm$\,7 as  derived by \citet{Birkin2021}, for SMGs with typical metallicity, star-formation rates and stellar masses \citep[e.g.,][]{Remy2014} and assuming  the F444W half-light radius provides an upper-limit on the total extent of the dust region, we obtain a lower-limit on the median $V$-band dust attenuation of \av\,=\,110\,$\pm$\,20. This is two orders of magnitude higher than that derived from the SED fitting of SMGs, with the median \texttt{\sc{magphys}} estimated value of \av\,=\,3.40\,$\pm$\,0.16. For the \av\ derived from \texttt{\sc{magphys}}, which constrains the dust obscuration of the visible stars, an energy balance calculation is assumed such that far-infrared emission is broadly consistent with the absorbed stellar light of the system \citep{Battisti2019}. This assumption assumes the UV/optical and far-infrared emission are co-spatial and well mixed however, inaccuracies can arise as detectable optical emission does not encode information about the ongoing obscured star formation of the galaxy \citep[e.g.,][]{Simpson2017,Buat2019,Haskell2023,Killi2024}\rev{, with previous studies identifying a 15\% uncertainty in the SFRs derived from {\sc{magphys}} for galaxies from the {\sc{eagle}} simulation \citep[e.g.,][]{Dud2020}}. In the derivation of \av\ from the dust mass, which defines the dust obscuration of the deepest regions of the clouds, we have estimated a Compton thick HI column where the dust is uniformly distributed in a smooth ``disk-like'' component as identified in the high-resolution ALMA studies of SMGs \citep[e.g.,][]{Hodge2016,Gullberg2019}. However, numerical studies suggest that the star dust geometry plays a crucial role in determining the attenuation curve of the galaxies \citep[e.g.,][]{Inoue2005,Sachdeva2022,Vijayan2024} as well as the metallicity of the interstellar medium \citep[e.g.,][]{Shivaei2020}.

\subsection{Dust Content and the interstellar medium of SMGs}

As noted earlier, while the properties derived assuming energy-balance are uncertain, they would be considerably less reliable if derived solely from the UV to near-infrared photometry and no energy-balance. With such high levels of inferred dust obscuration, and intense star-formation rates, the question remains what are underlying interstellar medium conditions that give rise to these extreme physical properties in the SMGs. To infer the properties of the interstellar medium we compare the surface density of star formation, ($\Sigma_{\rm SFR}$), to the gas surface density ($\Sigma_{\rm gas}$) i.e., the density of fuel for star formation. For the SMGs, we use the gas mass derived above, the star-formation rate from {\sc{magphys}} and the F444W half-light radius as a conservative estimate of the physical extent of the gas content. For the field sample, we use the {\sc{magphys}} derived dust masses with a median value of $\rm \log_{10}(M_d[M_{\odot}])$\,=\,7.7\,$\pm$\,0.1 \rev{and  $\rm \log_{10}(M_d[M_{\odot}])$\,=\,7.6\,$\pm$\,0.1 for the massive field}, which are consistent with the dust mass inferred from converting the $S_{870\um}$ limit, derived from stacks in S2CLS UDS 870\um\ map \citep{Geach2017}, to a dust mass using the $S_{870\um}$ to dust mass relation defined in \citet{Dud2020}. Adopting the metallicity-dependent dust-to-gas ratio from \citet{Tacconi2018} for massive star-forming galaxies and following the mass metallicity relation defined in \citet{Genzel2015}, we derive a median dust-to-gas ratio of $\delta$\,=\,163\,$\pm$\,32 which equates to a median gas mass of $\rm \log_{10}(M_g[M_{\odot}])$\,=\,9.8\,$\pm$\,0.1 for the field galaxies. \rev{Following the same procedure but for the massive field galaxies, we derive a median gas mass of $\rm \log_{10}(M_g[M_{\odot}])$\,=\,9.9\,$\pm$\,0.1 for the massive field sample.}

\begin{figure}
\centering
\includegraphics[width=\linewidth]{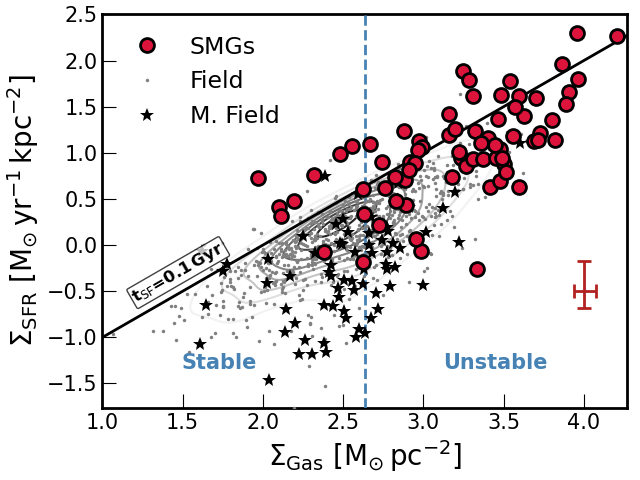}
 \caption{The relation between star-formation rate surface density and gas surface density for the field sample (grey points and contour) \rev{and massive field sample (black stars)} and SMGs (red circles). We indicate a constant star-formation time scale of 0.1\,Gyr with the black line. Adopting the typical rotation velocity and velocity dispersion from \citet{Birkin2023} derived for a sample of AS2UDS SMGs, we define the boundary between stable ($Q$\,$>$\,1) and unstable ($Q$\,$<$\,1) gas disks. The majority of the SMGs indicate unstable gas disks with a median Toomre $Q$ parameter of $Q$\,=\,0.47\,$\pm$\,0.06 while most of the field samples are stable or marginally stable.}
 \label{Fig:KS}
\end{figure}

Combining this gas mass estimate for the field samples with the {\sc{magphys}} derived star-formation rates and F444W half-light radius, in Figure \ref{Fig:KS} we show the relation between $\Sigma_{\rm SFR}$ and $\Sigma_{\rm gas}$ for both the field \rev{populations} and SMGs. Given the higher star-formation rates and dust masses of the SMGs (Figure \ref{Fig:SED_hist}) and their comparable sizes in the F444W band to the field  galaxies (Figure \ref{Fig:Morph_hist}), the SMGs have considerably higher star-formation rate surface density and gas surface density. However, both \rev{the lower mass} field galaxies and SMGs have comparable time scales for star formation, lying close to the 0.1\,Gyr star formation time scale. \rev{The massive field sample has lower $\Sigma_{\rm SFR}$, due to their lower specific star formation rates (Figure \ref{Fig:Morph_hist}).}

By estimating the Toomre Q parameter \citep{Toomre1964} we can place constraints on the interstellar medium properties of the galaxies, in particular on the stability of the gas disk. The Toomre Q is defined as,
\begin{equation}
     Q\,=\,\frac{\sigma \kappa}{\pi G \Sigma_{\rm gas}}
\end{equation}
where $\sigma$ is the radial velocity dispersion, $\kappa$ is the epicylic frequency, for which we adopt $\kappa$\,=\,$\rm \sqrt{2}V/R$ appropriate for a galaxy with a flat rotation curve and $\rm \Sigma_{gas}$ is the gas surface density. For the rotation velocity and velocity dispersion, we adopt the median values from \citet{Birkin2023}, for a sample 31 SMGs in the AS2UDS survey, with comparable median redshift and stellar mass to our sample ($z$\,$\sim$\,1.3\,--\,2.6, $\rm \log_{10}(M_{*}[M_{\odot}])$\,=\,11.11\,$\pm$\,0.06). \citet{Birkin2023} derive a median rotation velocity and velocity dispersion of $V_{\rm circ}$\,=\,230\,$\pm$\,20 km\,s$^{-1}$ and $\sigma$\,=\,87$\pm$\,6 km\,s$^{-1}$.

Adopting the F444W half-light radius as an estimate of the extent of the gas disk, we can derive the gas-surface density required for a quasi-stable gas disk, identifying a threshold value of $\log_{10}(\Sigma_{\rm gas}$)\,=\,2.64\,$\pm$\,0.05 $\rm M_{\odot}\,pc^{-2}$. This implies the gas disk is unstable to collapse ($Q$\,$<$\,1) for gas-surface densities higher than this value. We estimate that the majority of the SMGs lie in this region of ``instability'' with a median Toomre $Q$ parameter of $Q$\,=\,0.47\,$\pm$\,0.06, while the bulk of the field samples are stable ($Q$\,$>$\,1) or marginally stable ($Q$\,$\simeq$\,1).

To test the sensitivity of these conclusions on our assumptions, we note that given the F444W half-light radius provides a lower limit on the extent of the gas disk. Thus to generate a quasi-stable interstellar medium in the SMGs, we require a larger rotation velocity or velocity dispersion. If the gas disk is required to be quasi-stable i.e Q\,$\geq$\,1, we derive a lower limit on the radial velocity dispersion needed. Adopting the rotation velocity from \citet{Birkin2023}, we derive a median velocity dispersion for stability of $\sigma_{\rm stable}$\,$\geq$\,200\,$\pm$\,30 km\,s$^{-1}$ which is significantly larger than that derived for ionised gas in previous studies of high redshift SMGs or U/LIRGS \citep[e.g.][]{Hogan2021,Birkin2023,Amvrosiadis2024}, suggesting that the gaseous disks of the SMGs in our sample are very likely unstable to gravitational collapse.

\subsection{Implications for the nature of SMGs}

The SMG and field samples employed in our analysis share several characteristics. By construction, they have similar number densities as a function of redshift \rev{with the larger field sample matched in specific star formation rate over $z$\,$\simeq$\,1\,--\,5 and the massive field sample matched in stellar mass over $z$\,$\simeq$\,1\,--\,3}. Furthermore, we have demonstrated they have similar rates of mergers and signatures of disturbance, as well as comparable rest-frame near-infrared axis ratios and (mass normalised) stellar continuum (F444W) sizes. This suggests that the major structural components in both populations comprise (randomly orientated) disks that broadly follow the size-mass relation for disk galaxies at $z$\,$\sim$\,2\,--\,3. The similarity in the merger fractions for the SMGs and field is consistent with the theoretical investigation of SMGs in the EAGLE simulation by \citet{Mcalpine2019}, who concluded that there was no difference in merger statistics of SMGs and the general population in the simulations, instead, most massive galaxies at these redshifts were undergoing minor or major mergers.

However, there are several distinctions between the SMGs and the less-active field samples. The dust-mass-selected SMGs have much higher \av\ and asymmetry, and they show stronger variations in their morphologies at bluer wavelengths, indicating the presence of structured and centrally concentrated dust.  Moreover, \rev{both field samples have} an average S\'ersic index of $n$\,$\gtrsim$\,2, indicating the potential presence of a bulge component in these systems. While the SMGs have consistently lower S\'ersic index, with a median of $n$\,$\sim$\,1.1, indicating they are nearly pure exponential disks or that any bulge component if present is invisible at rest-frame $\sim$1\um. Either of these interpretations suggests that any bulge component in the SMGs is of low mass compared to the disk or is likely to be young and so only recently formed. 

We also infer more than an order of magnitude higher molecular gas surface densities for the SMGs than the field.  These differences suggest that one possible cause for a distinction between SMGs and less active galaxies arises from the less developed bulge components present in the SMGs (see \citealt{Mcalpine2019}). Consequently, their gas disks are both more massive and also more sensitive to developing low-order disk instabilities, including $m$\,=\,1 bars \citep[e.g.,][]{Marinova2007,pettitt2018}. These instabilities  can either be triggered by external perturbations from major or minor mergers or tidal interactions, or from secular instabilities in the dense gas disks.  These structures, as predicted by theoretical models \citep[e.g.,][]{Fragkoudi2021,Bland-Hawthorn2023} and identified in recent studies of individual high-redshift SMGs \citep[e.g.,][]{Hodge2019,Smail2023,Amvrosiadis2024} serve to funnel gas from the dense gas disk into the central regions of the galaxies, driving a central starburst that creates both the compact sub-millimetre emission we see and the corresponding highly obscured central regions of these systems.  

\rev{To assess the possible subsequent structured evolution of the SMGs we follow the prescription of \citet{Brennan2015}, that relates B/T and S\'ersic index. Adopting the F444W/870\um\,=\,2.7\,$\pm$\,0.4 as the ratio of the disc to bulge size, the SMGs at $z$\,=\,2.7\,$\pm$\,0.2 with a $n_{\rm F444W}$\,=\,1.1\,$\pm$\,0.1 have a bulge-to-total ratio of B/T\,=\,0.16\,$\pm$\,0.12. If the SMGs converted their gas mass (which we assume to be equivalent to 50\% of their stellar mass following  \citealt{Birkin2021}) to a bulge component, it would result in a B/T\,=\,0.49\,$\pm$\,0.18 which following \citet{Brennan2015} gives a S\'ersic index of $n$\,=\,2.2\,$\pm$\,0.8. This is consistent with to the S\'ersic index measure for the gas-poor bulge-strong massive field sample, which suggests that the field galaxies may have already experienced their ``SMG'' phase at an earlier epoch and the SMGs may evolve to a gas-poor bulge-strong massive but less active population population within the next $\simeq$\,0.1\,Gyr.}

\section{Conclusions}\label{Sec:Conc}

We present a multi-wavelength morphological analysis of a complete sample of ALMA-identified submillimetre-selected galaxies from the AS2UDS and AS2COSMOS surveys \citep{Stach2019,Simpson2020} with $>$4.5$\sigma$ ALMA 870\um\ detections and positions. Utilising public \jwst/NIRCam and MIRI imaging from the PRIMER survey, we build a sample of 80 SMGs, determining their multi-wavelength properties through an SED and morphological analysis. We further define two matched samples of less active and more typical $K$-band selected star-forming field galaxies in the UDS field and analyse the differences between the populations. Our main conclusions are:
  
\begin{itemize}
    
\item From visually inspecting the F277W/F356W/F444W colour images and F444W {\sc{galfit}} residual maps of the SMGs we identify 20\,$\pm$\,5\% as candidate major mergers, a further 40\,$\pm$\,10\% as potential minor mergers and the remaining 40\% are isolated undisturbed systems. We find no dependence of the visual classification on far-infrared luminosity, \rev{with similar statistics identified in the field samples, suggesting that the majority of SMGs are not ongoing major mergers, although mergers (major or minor) and disturbed appearances are common in both the SMGs and the less active field population at these redshifts.}

\item The SMG and field populations are more distinct in their detailed morphologies in terms of S\'ersic index ($n$) and Asymmetry ($A$) in the F444W band, with the SMGs having  lower S\'ersic indices and higher Asymmetry ($n_{\rm F444W}$\,=\,1.1\,$\pm$\,0.1, $A_{\rm F444W}$\,=\,0.13\,$\pm$\,0.02) compared to both the less massive field galaxies ($n_{\rm F444W}$\,=\,1.9\,$\pm$\,0.1, $A_{\rm F444W }$\,=\,0.08\,$\pm$\,0.01) \rev{or the more  and massive field galaxies ($n_{\rm F444W}$\,=\,2.8\,$\pm$\,0.2, $A_{\rm F444W }$\,=\,0.07\,$\pm$\,0.02)}.
  
\item A similar disparity between the SMGs and  typical star-forming galaxies  is seen in the variation of their morphologies as a function of wavelength. The size of the SMGs declines more rapidly at longer wavelengths than both field populations ($\delta (R_{\rm h}/R_{1\um})/\delta \lambda$\,=\,$-$0.60\,$\pm$\,0.09 versus $\delta (R_{\rm h}/R_{1\um})/\delta \lambda$\,=\,$-$0.15\,$\pm$\,0.07 \rev{for the low mass field sample or $\delta (R_{\rm h}/R_{1\um})/\delta \lambda$\,=\,$-$0.04\,$\pm$\,0.09 for the massive field sample) with the} SMGs exhibiting lower S\'ersic indices at all wavelengths.

\item We further determine the brightest and reddest regions of the SMGs in the NIRCam imaging correspond to the highest surface brightness emission at 870\,\um\ seen by ALMA, indicating a strong connection between the colour and the local dust columns in the galaxies on $\simeq$\,kpc scales.
  
\item To understand the physical origin of the difference in morphologies between the SMGs and the field galaxies, we assume a representative gas-to-dust ratio to define the relation between star-formation rate surface density and gas surface density, suggesting that the SMGs have significantly higher gas surface densities than field galaxies. The SMGs on average fall in the unstable ($Q$\,$<$\,1) regime with a median Toomre $Q$ parameter of $Q$\,=\,0.47\,$\pm$\,0.06, while the bulk of the field populations are stable or marginally stable ($Q$\,$\gtrsim$\,1). 
\end{itemize}

Our analysis indicates that  SMGs and typical star-forming galaxies appear morphologically distinct in the rest-frame optical due to the higher dust content of the SMGs, which preferentially influences the central regions of the galaxies. While at longer wavelengths, sampling the rest-frame near-infrared, the SMGs and less active field galaxies show more similar mass-normalised sizes, although the SMGs exhibit lower S\'ersic parameters, suggesting they have weaker bulge components. Both the SMGs and field control samples have comparable rates of mergers and undisturbed galaxies, suggesting that mergers are not a unique driver of the activity in SMGs (as indicated by previous theoretical studies, \citealt{Mcalpine2019}).  Instead, we suggest that the higher gas surface densities and weaker bulge components (which may be linked through previous evolution), coupled with the perturbations caused by either major or  minor mergers, as well as secular processes, lead to correspondingly more intense star-formation activity in the SMGs, compared to the field.  Thus the defining characteristic of SMGs maybe their massive and gas-rich nature, coupled with relatively underdeveloped bulge components and correspondingly low black hole masses (see \citealt{Mcalpine2019}).

\begin{acknowledgements}
We would like to thank the PRIMER team for designing and executing the observations upon which this work is based. The observations analysed in this work are made with the NASA/ESA/CSA James Webb Space Telescope (DOI: \hyperlink{https://archive.stsci.edu/doi/resolve/resolve.html?doi=10.17909/z7p0-8481}{10.17909/z7p0-8481}). Certain aspects of the analysis in this paper were undertaken on SCUBA-2 data taken as part of Program ID MJLSC02. SG acknowledges financial support from the Villum Young Investigator grants 37440 and 13160 and the Cosmic Dawn Center (DAWN), funded by the Danish National Research Foundation (DNRF) under grant No. 140. IRS and AMS acknowledge STFC grant ST/X001075/1. BG acknowledges support from the Carlsberg Foundation Research Grant CF20-0644 ‘Physical pRoperties of the InterStellar Medium in Luminous Infrared Galaxies at High redshifT: PRISM- LIGHT’. TRG is grateful for support from the Carlsberg Foundation via grant No.~CF20-0534. YM acknowledges support of JSPS KAKENHI Grant Numbers JP17KK0098, JP22H01273 and JP23K22544. C.-C.C. acknowledges support from the National Science and Technology Council of Taiwan (NSTC 111-2112M-001-045MY3), as well as Academia Sinica through the Career Development Award (AS-CDA-112-M02). This work made use of the following open-source software: Astropy \citep{2013A&A...558A..33A,2018AJ....156..123A}, Photutils \citep{Phot2022}, Source Extractor \citep{Bertin1996}, SEP \citep{Barbary2016}, Eazy-py \citep{Brammer2021}, GriZli \citep{Brammer2022},
GalfitM \citep{Haubler2013}, Statmorph \citep{statmorph2019}, Topcat \citep{Topcat2005}.

Cloud-based data processing and file storage for this work is provided by the AWS Cloud Credits for Research program. The data products presented herein were retrieved from the Dawn \jwst\ Archive (DJA). DJA is an initiative of the Cosmic Dawn Center, which is funded by the Danish National Research Foundation under grant No. 140
\end{acknowledgements}

%-------------------------------------------------------------------
\bibliography{master}{}
\bibliographystyle{mnras}

\hypertarget{orcids}{\section*{ORCIDs}}

Steven Gillman \orcid{0000-0001-9885-4589}\\
Ian Smail \orcid{0000-0003-3037-257X}\\
Bitten Gullberg \orcid{0000-0002-4671-3036},\\
A. M. Swinbank \orcid{0000-0003-1192-5837}\\
Aswin P. Vijayan \orcid{0000-0002-1905-4194},\\
Gabe Brammer \orcid{0000-0003-2680-005X}\\
Thomas R. Greve \orcid{0000-0002-2554-1837}\\
Omar Almaini \orcid{0000-0001-9328-3991}\\
Malte Brinch \orcid{0000-0002-0245-6365}\\
Chian-Chou Chen \orcid{0000-0002-3805-0789}\\
Wei-Hao Wang \orcid{0000-0003-2588-1265}\\
Yuichi Matsuda \orcid{0000-0003-1747-2891}\\
Fabian Walter \orcid{h0000-0003-4793-7880}\\
Paul P. van der Werf \orcid{0000-0001-5434-5942}\\

\FloatBarrier
\begin{appendix}
\onecolumn

%SMG table
\begin{table*}
\section{SMG properties}\label{App:Table}
\caption{Cumulative index for filter coverage in \hst\ and \jwst\ observations} 
\begin{tabular}{cccccccccccc}
\hline
Index &1 &  2 & 4 & 8 & 16 & 32 & 64 & 128 & 256 & 512 & 1024 \\
\hline
\hst\ & F275W & F336W & F435W & F475W & F606W & F814W & F850LP & F105W & F125W & F140W & F160W \\
\jwst\ & F090W & F115W & F150W & F200W & F277W & F356W & F410M & F444W & F770W & F1800W \\
\hline
\label{Table:index}
\end{tabular}
\end{table*}

\subsubsection*{Table A.2:} Table summarising the AS2UDS and AS2COSMOS SMGs identified in the PRIMER survey. The table is only available in electronic form at the CDS webpage\footnote{\url{http://cdsweb.u-strasbg.fr/cgi-bin/qcat?J/A+A/}} or via anonymous ftp to cdsarc.u-strasbg.fr (130.79.128.5). 

\section*{Appendix B: SEDs}\label{App:SEDS}
A full version of Appendix B is available in electronic format on the ZENODO platform\footnote{\url{https://zenodo.org/records/13805467}}.

\end{appendix}
\end{document}